\documentstyle[eqsecnum,prd,aps,epsf]{revtex}
\draft
\begin{document}
\thispagestyle{empty}
{\baselineskip0pt
\leftline{\baselineskip16pt\sl\vbox to0pt{\hbox
{\it Department of Physics}
               \hbox{\it Waseda University}\vss}}
\rightline{\baselineskip16pt\rm\vbox to20pt{\hbox{WU-AP/117/00}
            \hbox{\today} 
\vss}}%
}
\vskip1cm
\begin{center}{\large \bf
Convergence to a self-similar solution in general relativistic gravitational collapse}
\end{center}
\vskip1cm
\begin{center}
 {\large 
Tomohiro Harada
\footnote{Email: harada@gravity.phys.waseda.ac.jp}
and Hideki Maeda
\footnote{Email: hideki@gravity.phys.waseda.ac.jp}.\\
{\em Department of Physics,~Waseda University, Shinjuku, 
Tokyo 169-8555, Japan}}
\end{center}

\begin{abstract}
We study the spherical collapse of a perfect fluid 
with an equation of state $P=k\rho$
by full general relativistic numerical simulations.
For $0<k\alt 0.036$, it has been known that there exists
a general relativistic counterpart
of the Larson-Penston self-similar Newtonian solution.
The numerical simulations strongly suggest that,
in the neighborhood of the center,
generic collapse
converges to this solution in an approach to a singularity
and that
self-similar solutions other than this solution,
including a ``critical solution'' in the 
black hole critical behavior, are relevant
only when the parameters which parametrize initial data 
are fine-tuned.
This result is supported by a mode analysis on the 
pertinent self-similar solutions.
Since a naked singularity forms 
in the general relativistic Larson-Penston solution
for $0<k\alt0.0105$,
this will be the most serious known counterexample against 
cosmic censorship.
It also provides strong evidence for the self-similarity hypothesis
in general relativistic gravitational collapse.
The direct consequence is that 
critical phenomena will be observed in the collapse of isothermal gas
in Newton gravity, and the critical exponent $\gamma$ 
will be given by $\gamma\approx 0.11$, though the order parameter
cannot be the black hole mass.
\end{abstract}

\pacs{PACS numbers: 04.20.Dw, 04.25.Dm, 04.40.Nr}

\section{introduction}
There is no characteristic scale in general relativity
as well as in Newton gravity.
A set of field equations is invariant by scale transformations
if we assume appropriate matter fields.
It implies the existence of scale-invariant solutions 
to the field equations.
Such solutions are called self-similar solutions, which are defined
by the existence of a homothetic Killing vector field.
Although the self-similar solutions are only special solutions of 
Einstein equations, it often has been supposed
that these solutions play an important role 
in situations where gravity is an essential 
ingredient in a spherically symmetric system 
(for example, Carr~\cite{carr1999}).
Such an assumption can be called the 
{\em self-similarity hypothesis}.

A spherically symmetric self-similar 
system of a perfect fluid 
has been widely researched.
Self-similar solutions in Newton gravity have been researched
in an effort to obtain simple and realistic solutions 
of gravitational collapse~\cite{penston1969,larson1969,shu1977,hunter1977}.
In particular, the Larson-Penston solution,
which is one of the self-similar solutions,
is believed 
to describe the central part of generic spherical collapse
of isothermal gas. 
Recent numerical simulations and results of mode analyses
strongly support this proposition~\cite{ti1999,hn1997,hm2000a,hm2000b}.
In general relativity, a spherically symmetric 
self-similar system was discussed
in various situations, such as cosmological voids, 
gravitational collapse, 
primordial black holes~\cite{ch1974,bh1978a,bh1978b}, and so on. 
The detailed structure of 
self-similar collapse solutions were analyzed~\cite{op1990}.
The discovery of the black hole critical behavior
shed light on the importance of a self-similar solution
as a critical solution~\cite{choptuik1993,ec1994}.
Several very recent works have been done
in complete classification of self-similar 
solutions~\cite{gnu1998a,gnu1998b,cc1999,carr2000,cc2000}.

In the context of the black hole critical behavior,
the self-similar critical solution is not an attractor.
A renormalization group approach
showed that the critical
solution has only one repulsive mode~\cite{kha1995,maison1996}.
The critical exponent which appears in the scaling law of the 
formed black hole mass
is equal to the inverse of the eigenvalue of 
the repulsive mode for a perfect fluid case. 

In the context of cosmic censorship~\cite{penrose1969,penrose1979}, 
a spherical system of a pressureless fluid (dust) 
has been extensively examined since
it can be solved exactly. 
It has been shown that a naked singularity forms in 
generic spherical collapse of dust
from an analytic initial density 
profile~\cite{es1979,christodoulou1984}.
It is noted that this solution is not self-similar at all.
In the presence of pressure, 
self-similar solutions were investigated by Ori and 
Piran~\cite{op1990,op1987,op1988a}.
For an adiabatic equation of state $P=k\rho$ ($0<k\alt 0.4$),
they found a discrete set of self-similar solutions 
which are analytic both at the center and at the sonic point.
They discovered the general relativistic counterpart of 
the Larson-Penston solution for $0<k\alt 0.036$.
They observed that a naked singularity forms in this solution
for $0<k\alt 0.0105$.
They also observed that there exist analytic self-similar solutions 
with naked singularity for every $k$.
Harada~\cite{harada1998} showed the generic occurrence 
of naked singularity in spherical 
collapse of a perfect fluid for $0<k\alt 0.01$  
by numerical simulations using the code based on the
Hernandez-Misner formulation without 
the ansatz of self-similarity.

The aim of this paper is to examine the 
validity of the self-similarity hypothesis
for a spherical system of a perfect fluid 
and to understand the relation of the 
self-similarity hypothesis, critical behavior
and cosmic censorship.
In this paper, we adopt the geometrized unit.
\section{Basic Equations}
\subsection{Einstein equation}
We adopt the comoving coordinates.
The line element in a spherically symmetric spacetime is given by
\begin{equation}
ds^{2}=-e^{\sigma(t,r)}dt^{2}+e^{\omega(t,r)}dr^{2}+R^{2}(t,r)(d\theta^{2}
+\sin^{2}\theta d\phi^{2}).
\end{equation}
It is noted that the comoving coordinates may be able to cover
even inside of the apparent horizon.
We consider a perfect fluid
\begin{equation}
T^{\mu\nu}=(\rho+P)u^{\mu}u^{\nu}+Pg^{\mu\nu}.
\end{equation}
Then the Einstein equations and the equations of motion for 
the matter are reduced to the following simple form:
\begin{eqnarray}
\frac{\partial m}{\partial r}&=&4\pi R^{2}\rho\frac{\partial R}
{\partial r}, 
\label{eq:dmdr}\\
\frac{\partial m}{\partial t}&=&-4\pi R^{2} P\frac{\partial R}
{\partial t}, \\
\frac{\partial \sigma}{\partial r}&=&-\frac{2}{\rho+P}
\frac{\partial P}{\partial r}, 
\label{eq:dsigmadr}\\
\frac{\partial \omega}{\partial t}&=&-\frac{2}{\rho+P}
\frac{\partial \rho}{\partial t}-\frac{4}{R}\frac{\partial R}
{\partial t}, 
\label{eq:domegadt}\\
m&=&\frac{R}{2}\left[1+e^{-\sigma}\left(\frac{\partial R}
{\partial t}\right)^{2}-e^{-\omega}\left(\frac{
\partial R}{\partial r}\right)^{2}\right],
\label{eq:m}
\end{eqnarray}
where $m(t,r)$ is called the Misner-Sharp mass.
We assume the following equation of state:
\begin{equation}
P=k\rho,
\end{equation}
where we assume that $0<k<1$.
For a barotropic equation of state,
the existence of self-similar solution demands
the above form.
Moreover, this equation of state will be valid for 
isothermal gas in Newton limit and for relativistically
high-density polytropes~\cite{op1990}.
We define the following dimensionless functions~\cite{bh1978b}:
\begin{eqnarray}
\eta&\equiv &8\pi r^{2}\rho,\\
S&\equiv&\frac{R}{r}, \\
M&\equiv &\frac{2m}{r}.
\end{eqnarray}
We also define the following zooming coordinates:
\begin{eqnarray}
\tau&\equiv & -\ln(-t), \\
z&\equiv & \frac{r}{-t}.
\end{eqnarray}

It is found that Eqs. (\ref{eq:dsigmadr}) and 
(\ref{eq:domegadt}) can be integrated as
\begin{eqnarray}
e^{\sigma}&=&a_{\sigma}(t)z^{\frac{4k}{1+k}}
\eta ^{-\frac{2k}{1+k}}, \\
e^{\omega}&=&a_{\omega}(r)\eta^{-\frac{2}{1+k}}S^{-4},
\end{eqnarray}
where $a_{\sigma}(t)$ and $a_{\omega}(r)$ are arbitrary functions.
This integrability is the advantage of the comoving coordinates.
These arbitrary functions correspond to
the freedom of rescaling
the time and radial coordinates as $\tilde{t}=\tilde{t}(t)$
and $\tilde{r}=\tilde{r}(t)$.
Hereafter, we restrict this gauge freedom by choosing
$a_{\sigma}=\mbox{const}$ and $a_{\omega}=1$.

Being transformed into the zooming coordinates,
the field equations are reduced to
\begin{eqnarray}
&&M+M^{\prime}=\eta S^{2}\left(S+S^{\prime}\right), \\
&&\dot{M}+M^{\prime}
=-k\eta S^{2}(\dot{S}+S^{\prime}), \\
&&\frac{M}{S}=1+a_{\sigma}^{-1}
\left(\eta z^{-2}\right)
^{\frac{2k}{1+k}} z^{2}(\dot{S}+S^{\prime})^2
-\eta^{\frac{2}{1+k}}S^{4}
(S+S^{\prime})^{2},
\label{eq:m/s}
\end{eqnarray}
where the derivatives are abbreviated as
\begin{eqnarray}
\dot&\equiv &\frac{\partial }{\partial \tau}, \\
^{\prime}&\equiv&\frac{\partial}{\partial \ln z}.
\end{eqnarray}
For later convenience, we define the quantity $y$
which is 
one third of the ratio of the ``average density''
of the region interior to $r$ to the local density at $r$, 
defined as
\begin{equation}
y\equiv\frac{M}{\eta S^{3}}.
\end{equation}
If we consider the regular center,
then it is found from Eq.~(\ref{eq:dmdr}) that
\begin{equation}
y=\frac{1}{3}
\end{equation}
at the regular center $z=+0$.
We can define two velocity functions $V_{z}$ and $V_{R}$. 
The $V_{z}$ is the velocity of the $z=\mbox{const}$ line relative to the fluid element,
which is written as
\begin{equation} 
V_{z}= -z e^{\frac{\omega-\sigma}{2}},
\end{equation}
while $V_{R}$ is the velocity of the $R=\mbox{const}$ line relative to the fluid element,
which is written as
\begin{equation}
V_{R}\equiv -e^{\frac{\omega-\sigma}{2}}
\frac{\left(\frac{\partial R}{\partial t}\right)}
{\left(\frac{\partial R}{\partial r}\right)}
=V_{z}\frac{\dot{S}+S^{\prime}}{S+S^{\prime}}.
\end{equation}

\subsection{Self-similar solutions}
For self-similar solutions, we assume that all dimensionless
quantities depend only
on $z$, i.e., 
\begin{eqnarray}
\eta&=&\eta(z), \\
S&=&S(z), \\
M&=&M(z), \\
\sigma&=&\sigma(z), \\
\omega&=&\omega(z).
\end{eqnarray}
The field equations are reduced to the following form:
\begin{eqnarray}
&&(\ln M)^{\prime}=\frac{k}{1+k}\left(y^{-1}-1
\right), 
\label{eq:dmdz}\\
&&(\ln S)^{\prime}=-\frac{1}{1+k}\left(1-y
\right), 
\label{eq:dsdz}\\
&&(1-y)^{2}V_{z}^{2}-(k+y)^{2}+(1+k)^{2}\eta^{-\frac{2}{1+k}}S^{-6}\left(1-\frac{M}{S}\right)
=0,
\label{eq:constraint}
\end{eqnarray}
where $V_{z}$ and $V_{R}$ are written as
\begin{equation}
V_{z}\equiv -a_{\sigma}^{-\frac{1}{2}}
z^{\frac{1-k}{1+k}}\eta^{-\frac{1-k}{1+k}}S^{-2}
\label{eq:Vz}
\end{equation}
and 
\begin{equation}
V_{R}=-V_{z}\frac{1-y}{k+y},
\end{equation}
respectively.
It is noted that, although the apparent form of these equations
does not seem to be an autonomous system,
the original system before we have performed explicit integrations
is, of course, an autonomous system. 
We can find that the fluid velocity, with respect to 
$R=\mbox{const}$, vanishes only if $V_{z}=0$ or $y=1$.
The above set of equations together 
with appropriate boundary conditions
is enough to determine the unknown functions
$M(z)$, $S(z)$ and $\eta(z)$.
However, in addition to the above equations, the following dependent 
equation is used:
\begin{equation} 
(\ln \eta)^{\prime}=2-\frac{4 y V_{z}^{2}
-(1+k)^2 S^{-4}\eta^{-\frac{1-k}{1+k}}}{2(V_{z}^{2}-k)}.
\end{equation}

Using the fact that $y=1/3$ is satisfied at $z=+0$,
from Eqs.(\ref{eq:dmdz})--(\ref{eq:constraint}), 
we find the behavior of the solution around
the regular center $z=+0$ as
\begin{eqnarray}
M&=&\frac{C_{k}}{3}(2D)^{\frac{k}{1+k}}z^{\frac{2k}{1+k}}\left[1
+O\left(z^{\frac{2(1+3k)}{3(1+k)}}\right)\right], \\
S&=&C_{k}^{1/3}(2D)^{-\frac{1}{3(1+k)}}z^{-\frac{2}{3(1+k)}}
\left[1+O\left(z^{\frac{2(1+3k)}{3(1+k)}}\right)\right], \\
\eta&=& 2D z^{2} \left[ 1+O\left(z^{\frac{2(1+3k)}{3(1+k)}}\right)\right],
\end{eqnarray}
and
\begin{eqnarray}
y&=&\frac{1}{3}
\left[ 1+O\left(z^{\frac{2(1+3k)}{3(1+k)}}\right)\right], 
\end{eqnarray}
where $C_{k}$ is a constant determined by $k$ as
\begin{equation}
C_{k}\equiv \frac{3(1+k)}{1+3k},
\end{equation}
and the parameter $D$ is defined as
\begin{equation}
D\equiv \frac{1}{2}\lim_{z\to+0}z^{-2}\eta =
4\pi \rho(t,0) t^{2}.
\label{eq:definitionD}
\end{equation}
Therefore, solutions that have regular centers 
are parametrized by only one parameter $D$.
Now that we find
\begin{equation}
e^{\sigma}=a_{\sigma}(2D)^{-\frac{2k}{1+k}}
\end{equation}
at the regular center, we let $e^{\sigma}$ 
be unity at the regular center 
by choosing the constant $a_{\sigma}$ as
\begin{equation}
a_{\sigma}=(2D)^{\frac{2k}{1+k}}.
\end{equation}
This gauge fixing gives the physical meaning 
to the parameter $D$.
Then the behavior of $V_{z}$ around the regular center is written as
\begin{eqnarray}
V_{z}&=&-C_{k}^{-2/3}(2D)^{-\frac{1}{3(1+k)}}
z^{\frac{1+3k}{3(1+k)}} 
\left[1+O\left(z^{\frac{2(1+3k)}{3(1+k)}}\right)\right].
\end{eqnarray}

The system of equations has an apparent singularity at a point $z=z_{sp}$
at which the relative velocity of $z=\mbox{const}$ world line with respect
to the fluid element is equal to the sound speed, i.e.,  
\begin{equation}
V_{z}^{2}=k.
\label{eq:sonic surface}
\end{equation}
Such a point is called a sonic point. 
The regularity requires the following condition
at the sonic point:
\begin{equation}
4yk-(1+k)^2\eta^{-\frac{1-k}{1+k}}S^{-4}=0.
\label{eq:sonic line}
\end{equation}
Every regular solution
must cross the sonic point, satisfying Eq.~(\ref{eq:sonic line}).   

Ori and Piran~\cite{op1990} 
discovered the band structure
of solutions regular both at the center 
and at the sonic point.
In particular, there is only a discrete set of
solutions which are analytic at the sonic point.
Here, we give special attention to such solutions.
One of such solutions is the flat Friedmann (FF) solution.
There are another types of analytic solutions which are called
the ``black hole'' type solutions,
in which a massive singularity forms at $t=0$ and
after that the mass of the singularity grows linearly with $t$,
and the ``repulsive'' type solutions, 
in which the central singularity which forms at $t=0$ disappears
instantaneously and the cloud begins to expand at $t=0$.
The solutions are characterized by the number of oscillations 
in the velocity field $V_{R}$.

Then we consider the behavior of the analytic similarity solutions.
For the FF solution, it is found that
\begin{equation}
D=\frac{2}{3}\frac{1}{(1+k)^2}.
\end{equation}
The solution is written as
\begin{eqnarray}	
M&=&\frac{C_{k}}{3}\left(\frac{4}{3(1+k)^2}\right)
^{\frac{k}{1+k}}z^{\frac{2k}{1+k}}, \\
S&=&C_{k}^{1/3}\left(\frac{4}{3(1+k)^2}\right)
^{-\frac{1}{3(1+k)}}z^{-\frac{2}{3(1+k)}}, \\
\eta&=&2Dz^2, \\
y&=&\frac{1}{3},
\end{eqnarray}
and
\begin{equation}
V_{z}=-C_{k}^{-\frac{2}{3}}(2D)^{-\frac{1}{3(1+k)}}
z^{\frac{1+3k}{3(1+k)}}.
\end{equation} 
For the FF, the big crunch occurs at $t=0$,
i.e., the singularity occurs at the same time everywhere.

For $k\alt 0.036$,
there exists a pure collapse solution.
It tends to the Larson-Penston solution in 
Newton gravity in the limit $k\to 0$. 
Hereafter we call this solution the general relativistic
Larson-Penston (GRLP) solution.
For this solution, we find that the value of the 
parameter $D$ is given by
\begin{equation}
D\approx 1.439 
\end{equation}
for $k=0.01$. 
This coincides with the result of 
Ori and Piran~\cite{op1990,op1987,op1988a}.

\begin{figure}[htbp]
\centerline{(a)
\epsfysize 11cm
\epsfxsize 11cm
\epsfbox{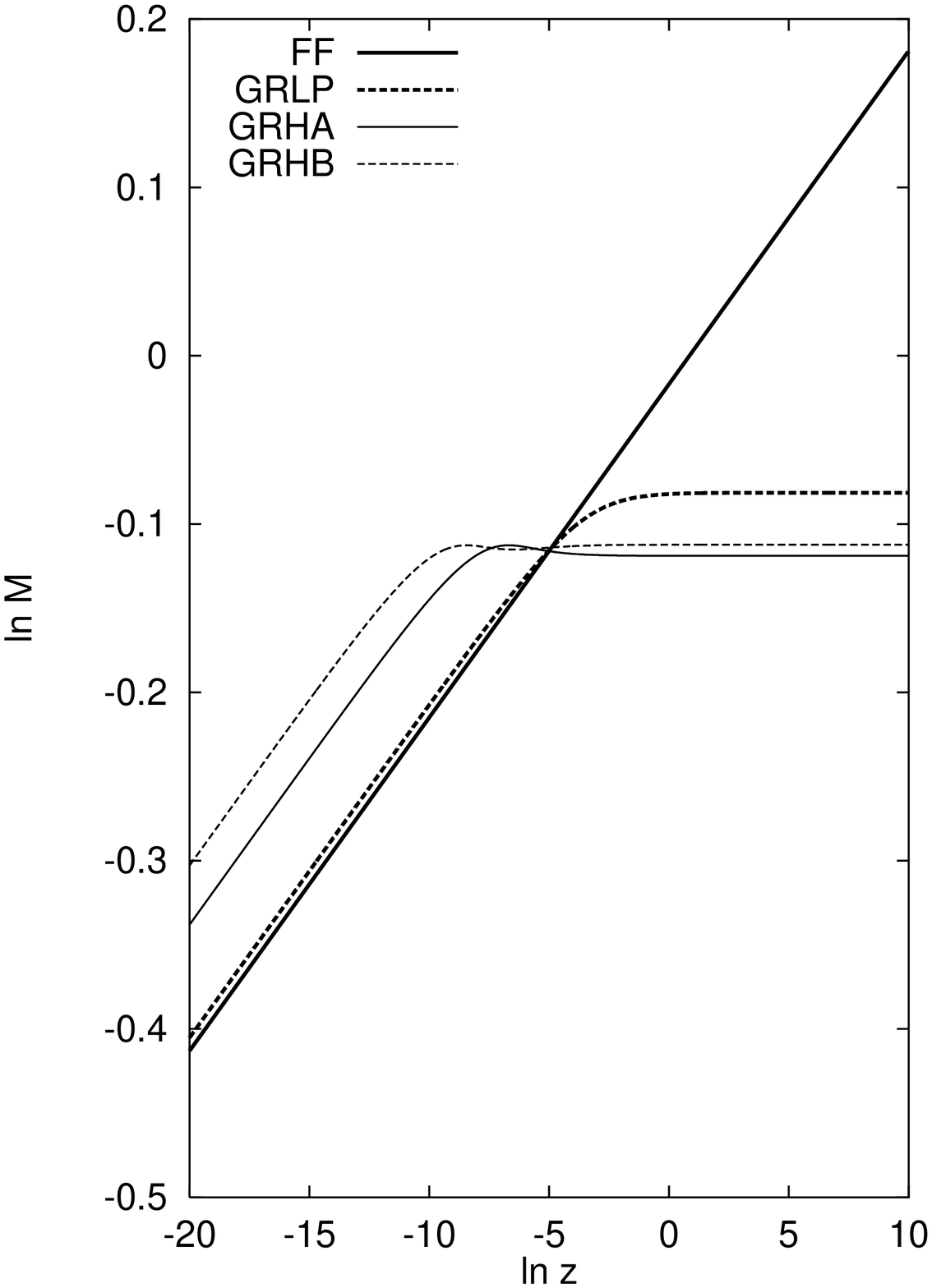}}
\centerline{(b)
\epsfysize 11cm
\epsfxsize 11cm
\epsfbox{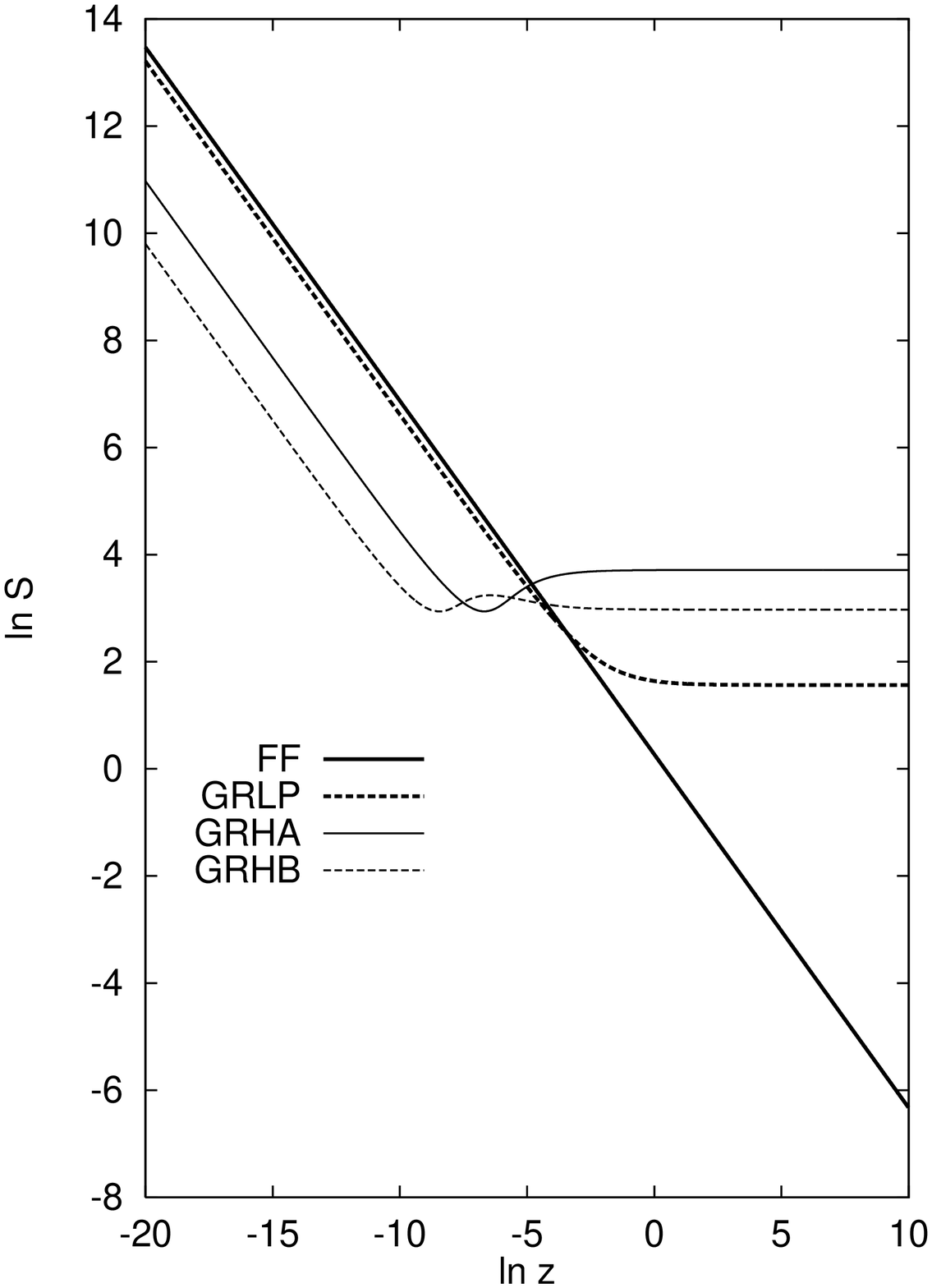}}
\centerline{(c)
\epsfysize 11cm
\epsfxsize 11cm
\epsfbox{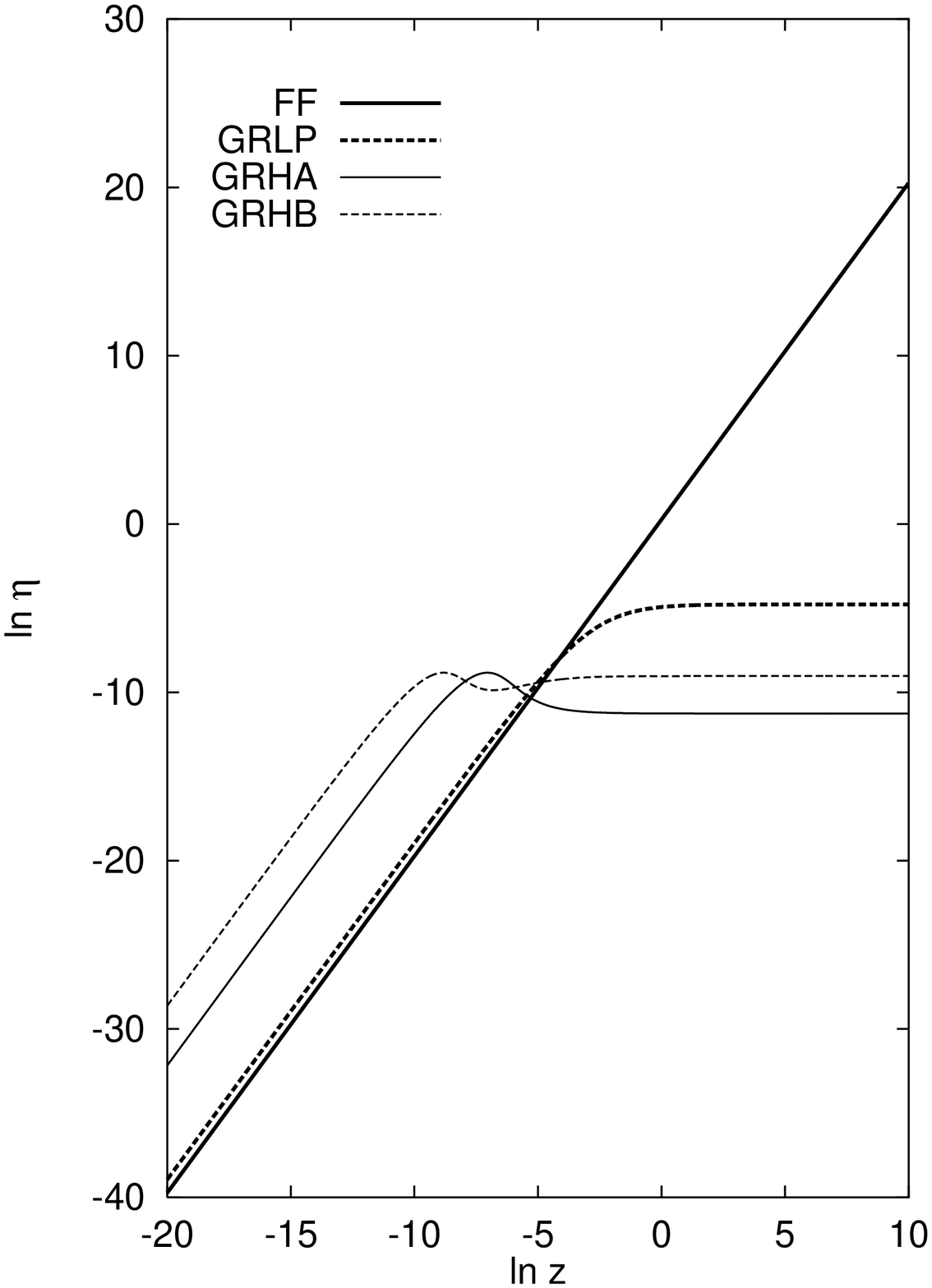}}
\centerline{(d)
\epsfysize 11cm
\epsfxsize 11cm
\epsfbox{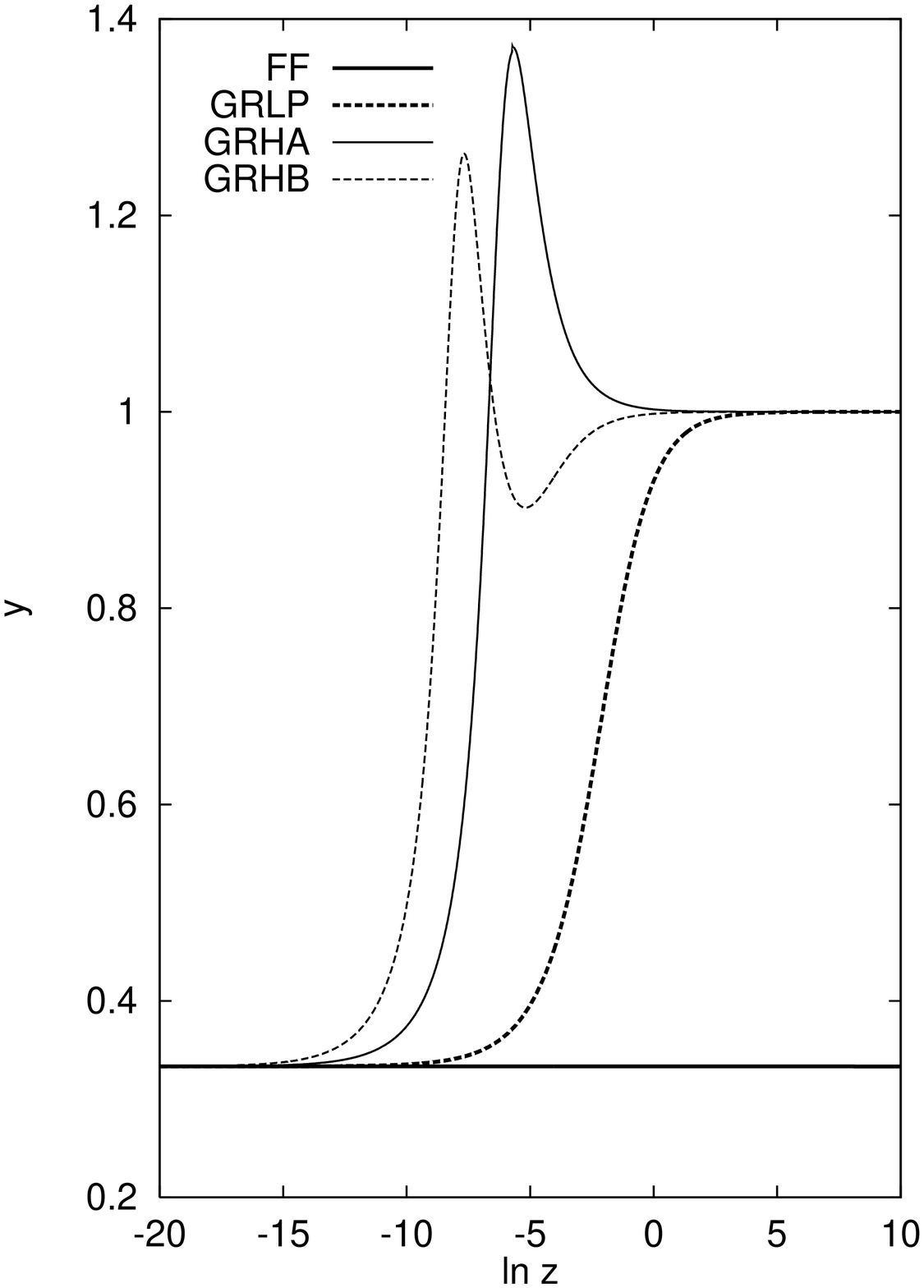}}
\centerline{(e)
\epsfysize 11cm
\epsfxsize 11cm
\epsfbox{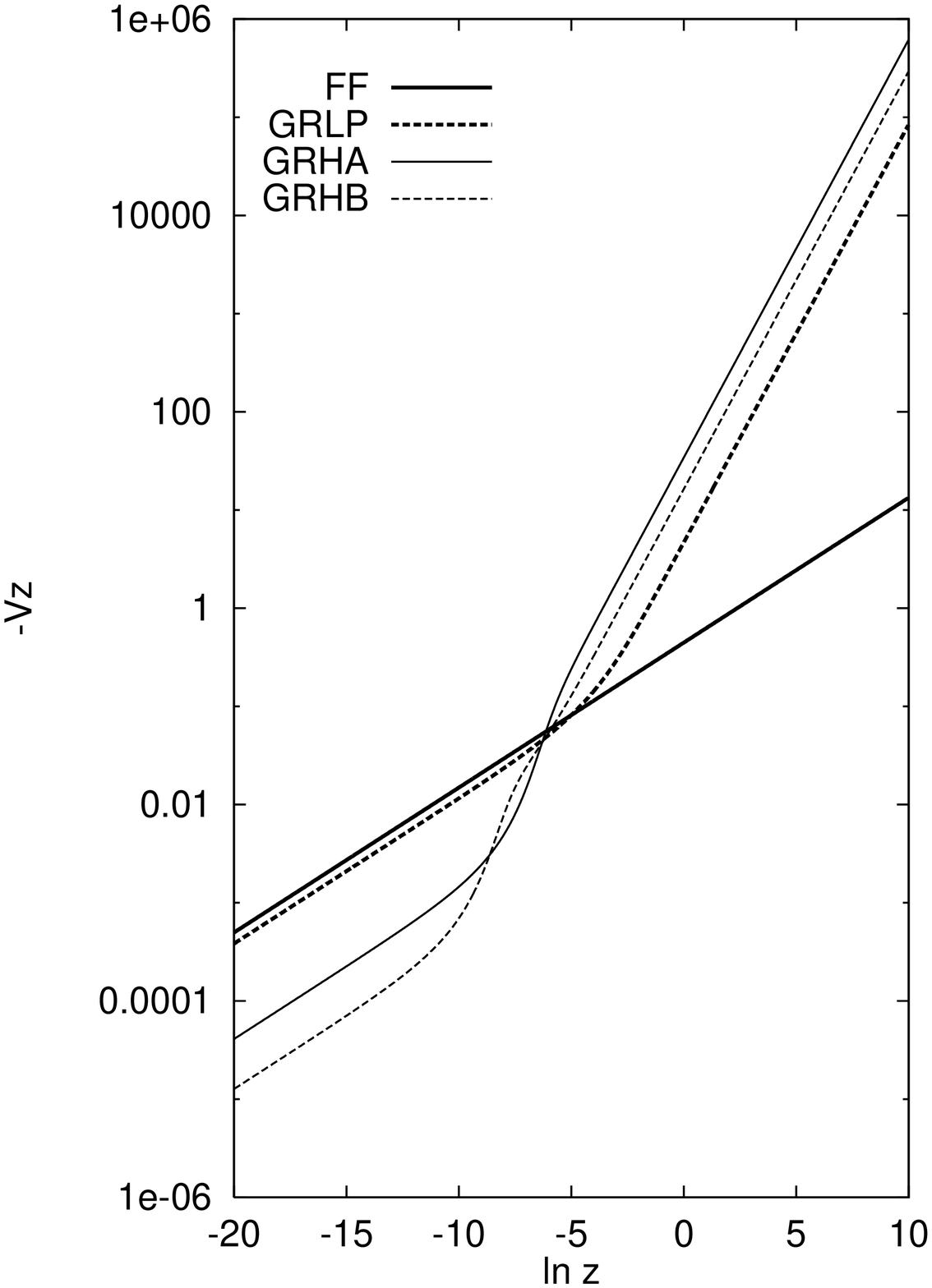}}
\centerline{(f)
\epsfysize 11cm
\epsfxsize 11cm
\epsfbox{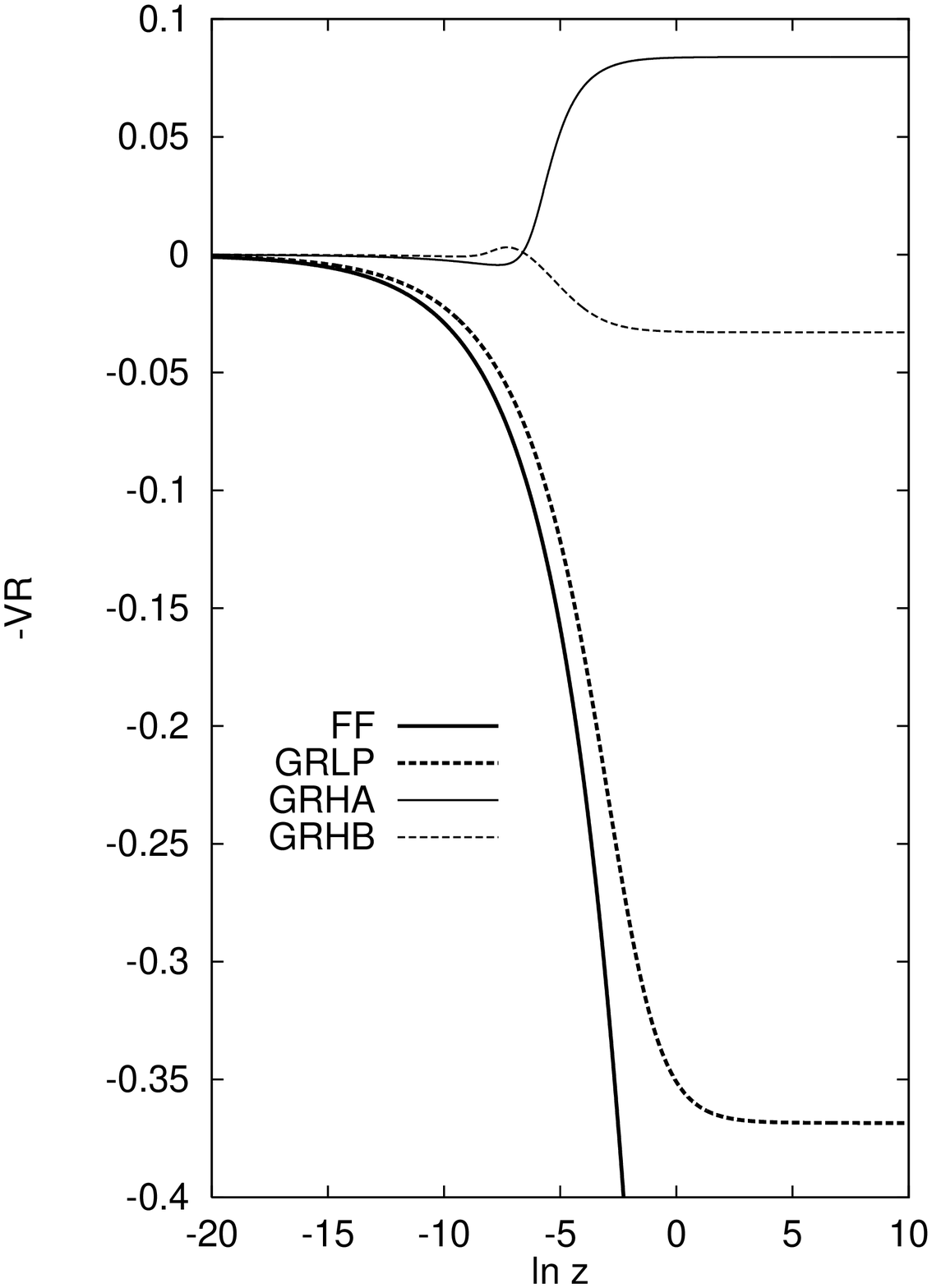}}
\centerline{(g)
\epsfysize 11cm
\epsfxsize 11cm
\epsfbox{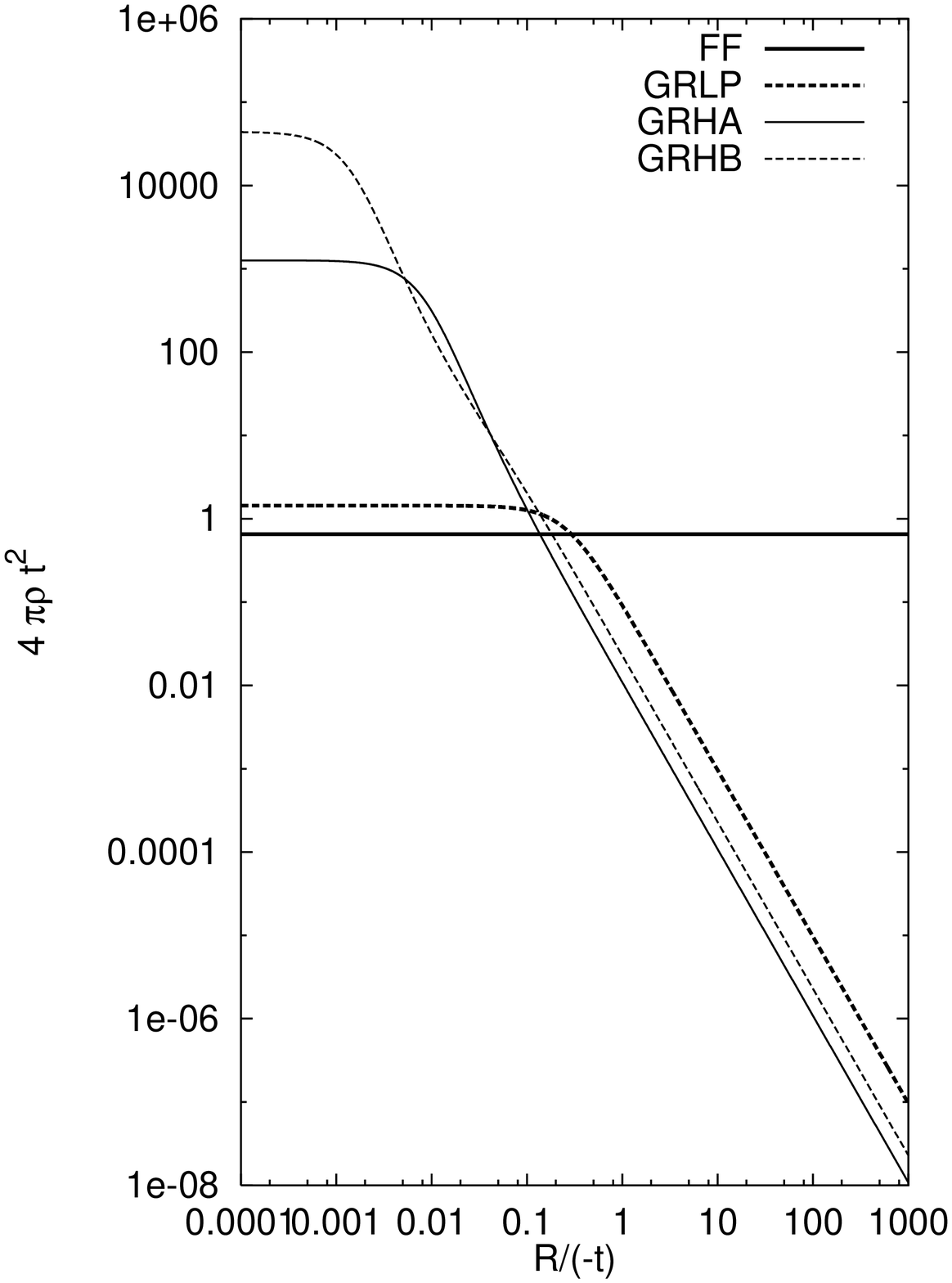}}
\caption{Self-similar solutions for $k=0.01$.
(a) $\ln M$, (b) $\ln S$, (c) $\ln\eta$, 
(d) $y$, (e) $-V_{z}$, and (f) $-V_{R}$ 
are plotted. In (g), the ordinate and abscissa are $4\pi \rho t^{2}$
and $R/(-t)$, respectively.}
\label{fg:sss}
\end{figure}

We have found another two analytic solutions.
These solutions are general relativistic counterparts 
of Newtonian self-similar solutions, Hunter (a) and (b)~\cite{hunter1977}.
We call these solutions GRHA and GRHB solutions, respectively.
These similarity solutions are displayed in 
Figs.~\ref{fg:sss}(a)--\ref{fg:sss}(g).

The FF is the only solution which 
has the big crunch singularity.
Unlike the FF,
the solution of
the black hole and repulsive types is
regular at $t=0$, except for at $r=0$. It implies that
the dimensionless physical quantities, such as
$M$, $S$ and $\eta$, are finite: i.e., 
\begin{eqnarray}
M&=&M_{\infty}, \\
S&=&S_{\infty}, \\
\eta&=&\eta_{\infty}=\frac{M_{\infty}}{S_{\infty}^{3}}, \\
y&=&1,
\end{eqnarray}
and
\begin{equation}
V_{R}=V_{R\infty},
\end{equation}
at $z=\pm\infty$.
The solution is black hole type if $V_{R\infty}>0$
and repulsive type if $V_{R\infty}<0$.
It is found that
the GRLP and GRHB are black hole type solutions while
the GRHA is a repulsive type one.
From the above equations, we find
\begin{equation}
\rho=\frac{\eta_{\infty}S_{\infty}^{2}}{8\pi R^{2}},
\end{equation}
at $t=0$.
The velocity function $V_{z}$ diverges as $z\to\infty$ as
\begin{equation}
V_{z}\approx -(2D)^{-\frac{k}{1+k}}\eta_{\infty}
^{-\frac{1-k}{1+k}}S_{\infty}^{-2}z^{\frac{1-k}{1+k}}.
\end{equation}  
The number of oscillations of $V_{R}$, which 
coincides with the number of zeroes of $(y-1)$
in the domain $0<z<\infty$,
is 0, 0, 1 and 2 for the FF, GRLP, GRHA,
and GRHB,
respectively.
The value of $D$ we have obtained for self-similar solutions
are summarized in Table~\ref{tb:D} for $k=0.001$, $0.008$, $0.01$, and $0.03$.

\section{numerical evidence for convergence to GRLP}
\label{sec:numerical evidence}
\subsection{Numerical simulations}
\label{subsec:numerical simulations}
In order to see the generic feature of
gravitational collapse,
we have numerically simulated the spherical collapse
of a perfect fluid.
We have numerically solved the full Einstein equations
(\ref{eq:dmdr})-(\ref{eq:m}) by the 
standard Misner-Sharp code without the self-similarity ansatz.
The finite difference equations have been given by the 
staggered-leapfrog scheme.
The distribution of grid points has been not homogeneous
but concentrated in the neighborhood of the center.
The total number of the grid points has been 10000.
See Harada~\cite{harada1998} for details of the numerical code
and references are therein.

For simplicity we display the results for 
time symmetric initial data.
It should be noted that we have confirmed that 
the results do not change so much
for several models in which initial data is not time symmetric.
As a set of initial data,
we have prepared both homogeneous and inhomogeneous 
balls of a perfect fluid which are momentarily 
static with vacuum external.
For the inhomogeneous models, the density profile has been given by 
\begin{equation}
\rho=\left\{\begin{array}{ll}
	\rho_{c,i}\left[1-\left(\displaystyle{\frac{R}{R_{s,i}}}\right)^{2}
	\right],&
	\quad 0\le R\le R_{s,i}, \\
        0, &
        \quad R_{s,i}<R.
        \end{array}\right.        
\end{equation}
Since we have confirmed that 
the results do not depend so much on 
the detailed form of the density profile,
the above functional form is considered to 
represent a typical situation.
The models which we have simulated 
are summarized in Table~\ref{tb:models},
where ${\cal M}$ is the Arnowitt-Deser-Misner
(ADM) mass of the ball.

\begin{figure}[htbp]
\centerline{(a)
\epsfysize 11cm
\epsfxsize 11cm
\epsfbox{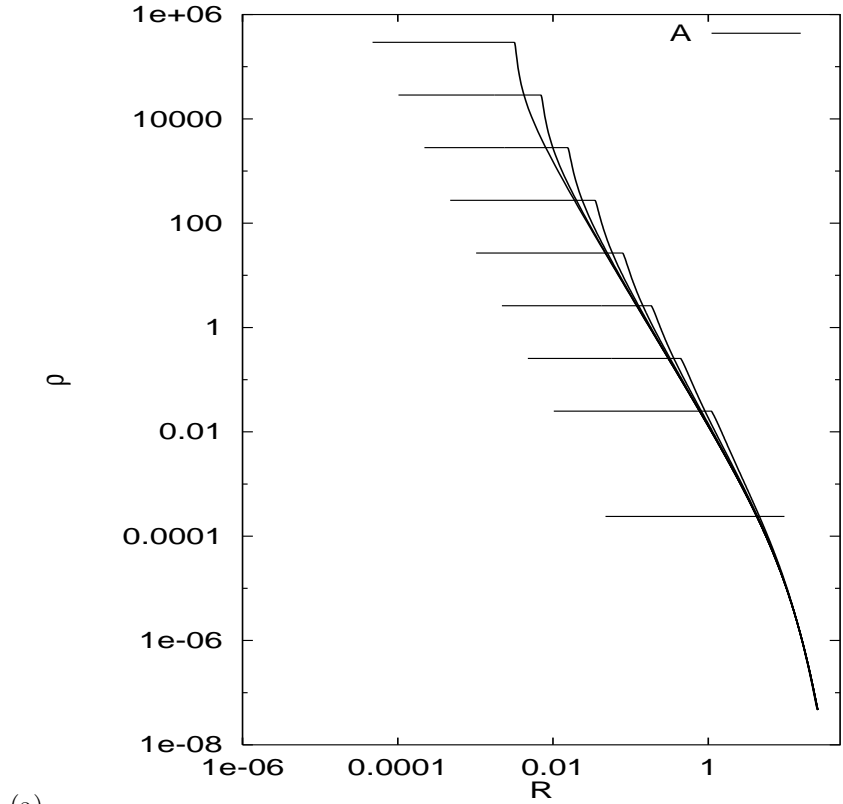}}
\centerline{(b)
\epsfysize 11cm
\epsfxsize 11cm
\epsfbox{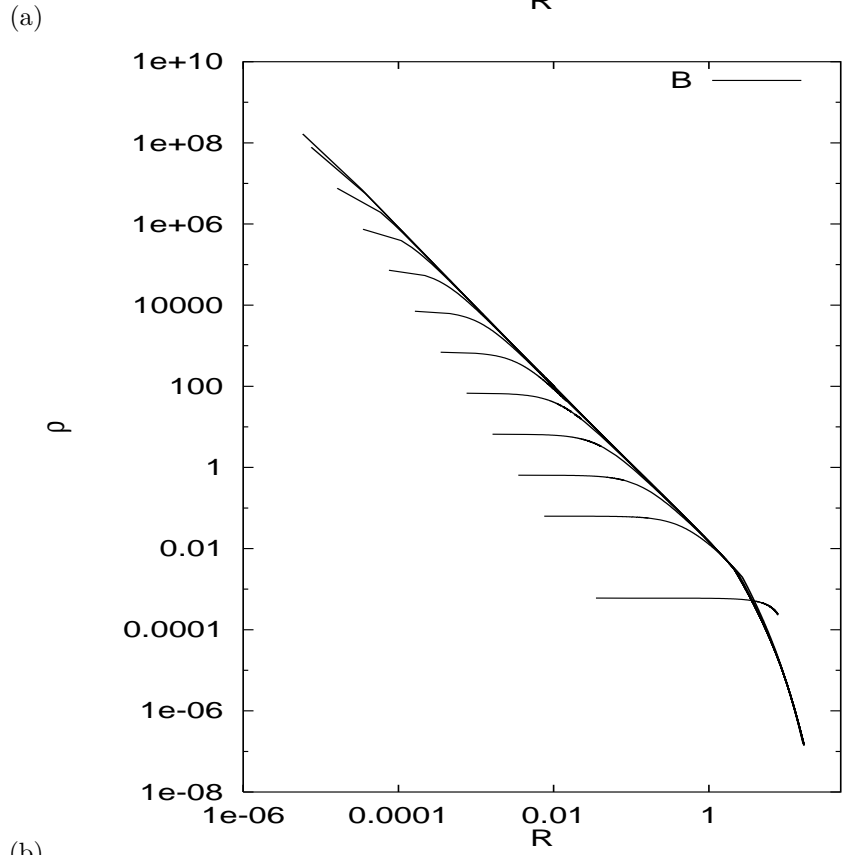}}
\caption{Time evolution of the density profile
for models (a) A and (b) B for $k=0.01$ are plotted. 
The ordinate and abscissa 
are the density $\rho$ and the circumferential radius $R$,
respectively. The unit is chosen so that the ADM mass ${\cal M}$ is unity.}
\label{fg:rd}
\end{figure}

\begin{figure}[htbp]
\centerline{(a)
\epsfysize 11cm
\epsfxsize 11cm
\epsfbox{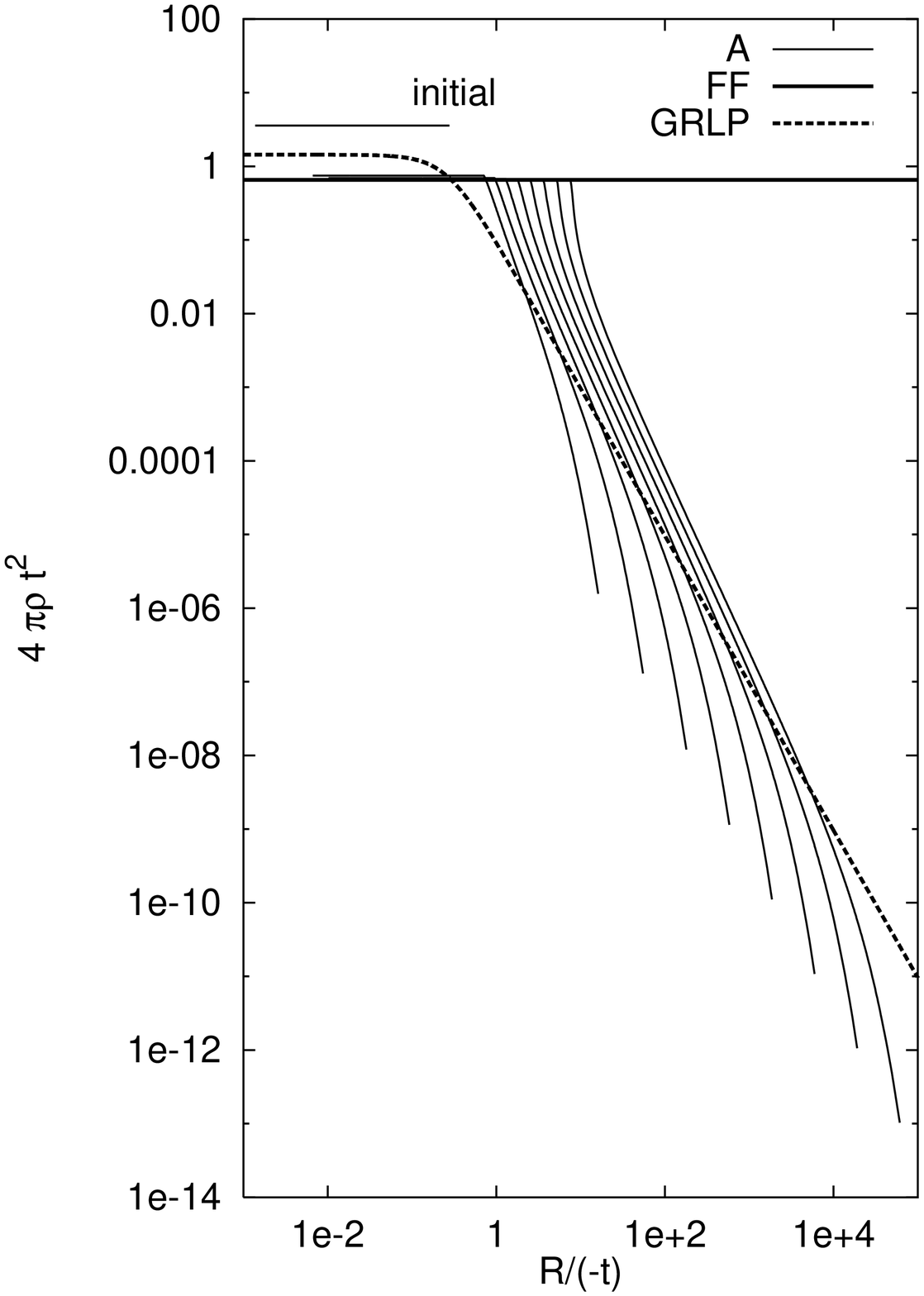}}
\centerline{(b)
\epsfysize 11cm
\epsfxsize 11cm
\epsfbox{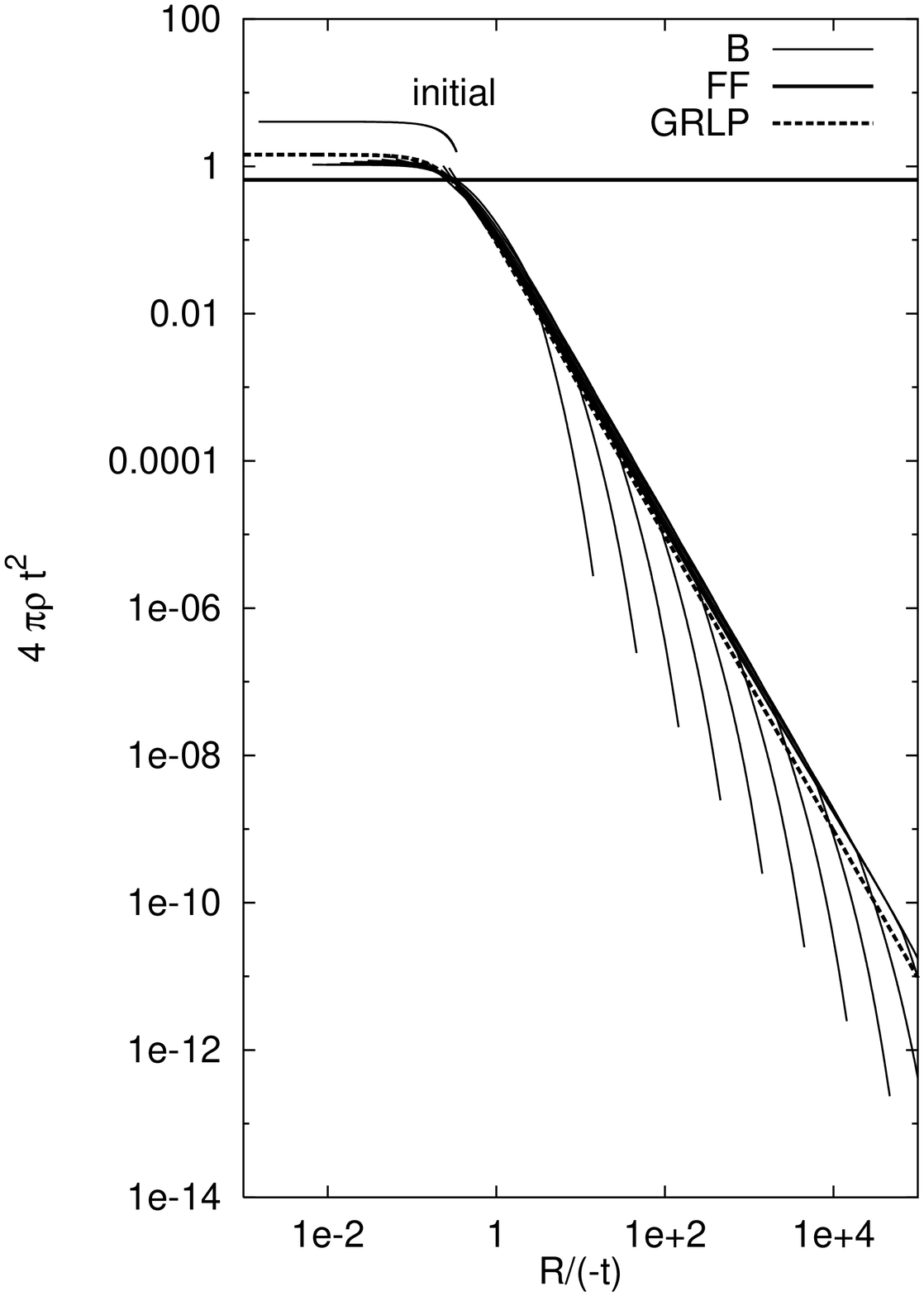}}
\caption{Time evolution of the density profile
for models (a) A and (b) B for $k=0.01$. 
The ordinate and abscissa are
$4\pi \rho t^{2}$ and $R/(-t)$, respectively.
For comparison, the FF and GRLP are also plotted.}
\label{fg:dimensionless}
\end{figure}

For $k=0.01$, we plot the time evolution of
the density profile of models A and B in 
Figs.~\ref{fg:rd}(a) and \ref{fg:rd}(b), respectively.
We can see that model B collapses
in a self-similar manner near the center
as the collapse proceeds.
In particular,
the density profile around the center 
tends to $\rho\propto R^{-2}$ in an approach to
the occurrence of singularity,
which is characteristic to the self-similar solutions
as we have seen.
In order to see more clearly that the collapse approaches 
the self-similar solution, 
we plot in Figs.~\ref{fg:dimensionless}(a) and 
\ref{fg:dimensionless}(b) 
the time evolution of the density profile
of models A and B for $k=0.01$, respectively. 
The ordinate and abscissa are 
dimensionless quantities
$4\pi \rho t^{2}$ and $R/(-t)$, respectively,
where $t$ is the proper time at the center and $t=0$ is chosen  
as the occurrence of singularity.
For comparison, we also plot the FF and GRLP in these figures
using the relations $4\pi\rho t^{2}=(1/2)\eta z^{-2}$
and $R/(-t)=S z$.
It is found that model B approaches the GRLP
while model A approaches the FF in an approach 
to the singularity.
As seen in Figs.~\ref{fg:sss}(b) and \ref{fg:sss}(e), 
$R/(-t)\simeq 0.26$ and $0.28$ at the sonic point 
for the FF and GRLP, respectively.
Therefore, from Figs.~\ref{fg:dimensionless}(a) and 
\ref{fg:dimensionless}(b),
it is found that the approach to the FF and GRLP is not only for the
subsonic region but also for the supersonic region.

\begin{figure}[htbp]
\centerline{(a)
\epsfysize 11cm
\epsfxsize 11cm
\epsfbox{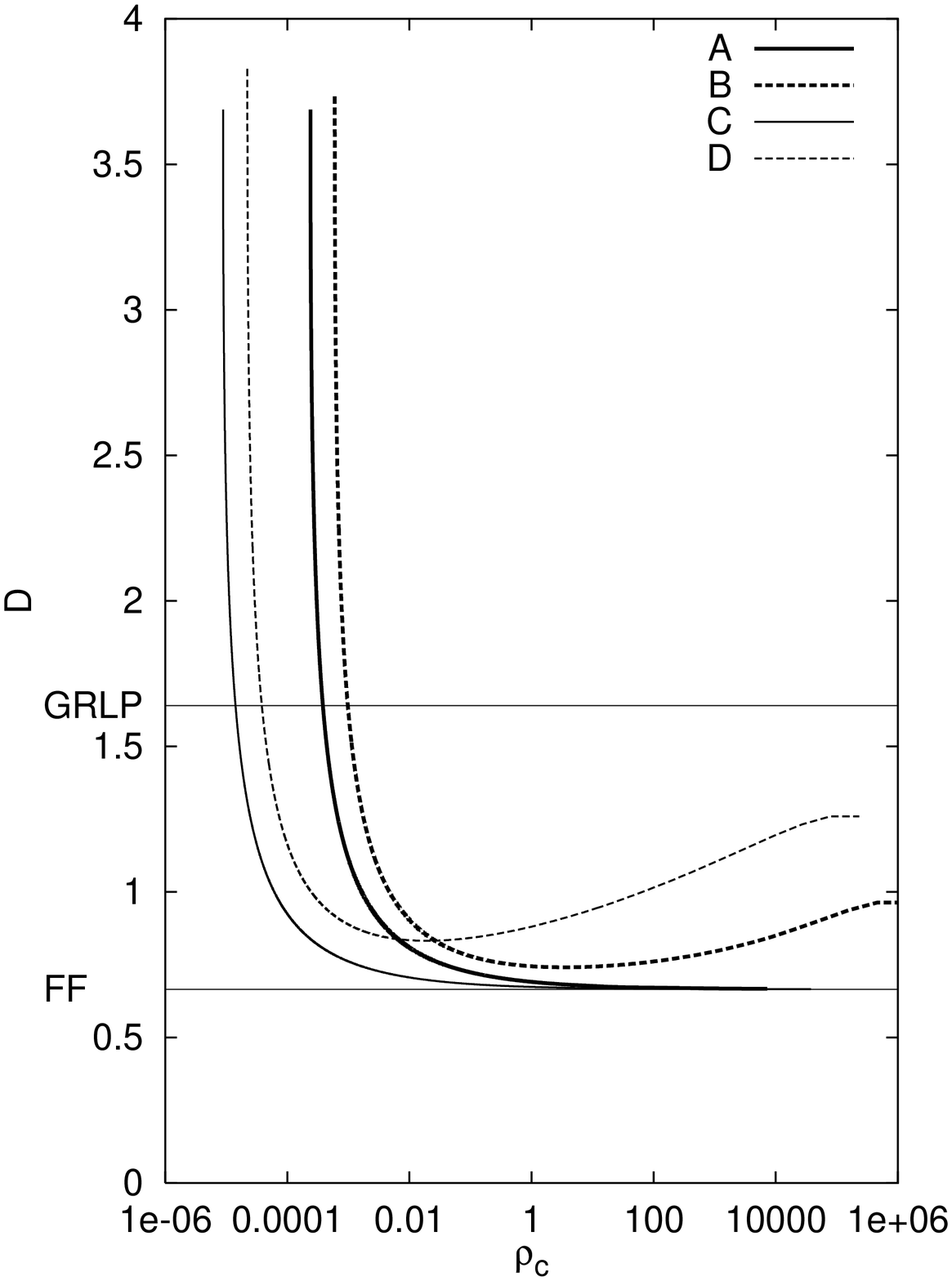}}
\centerline{(b)
\epsfysize 11cm
\epsfxsize 11cm
\epsfbox{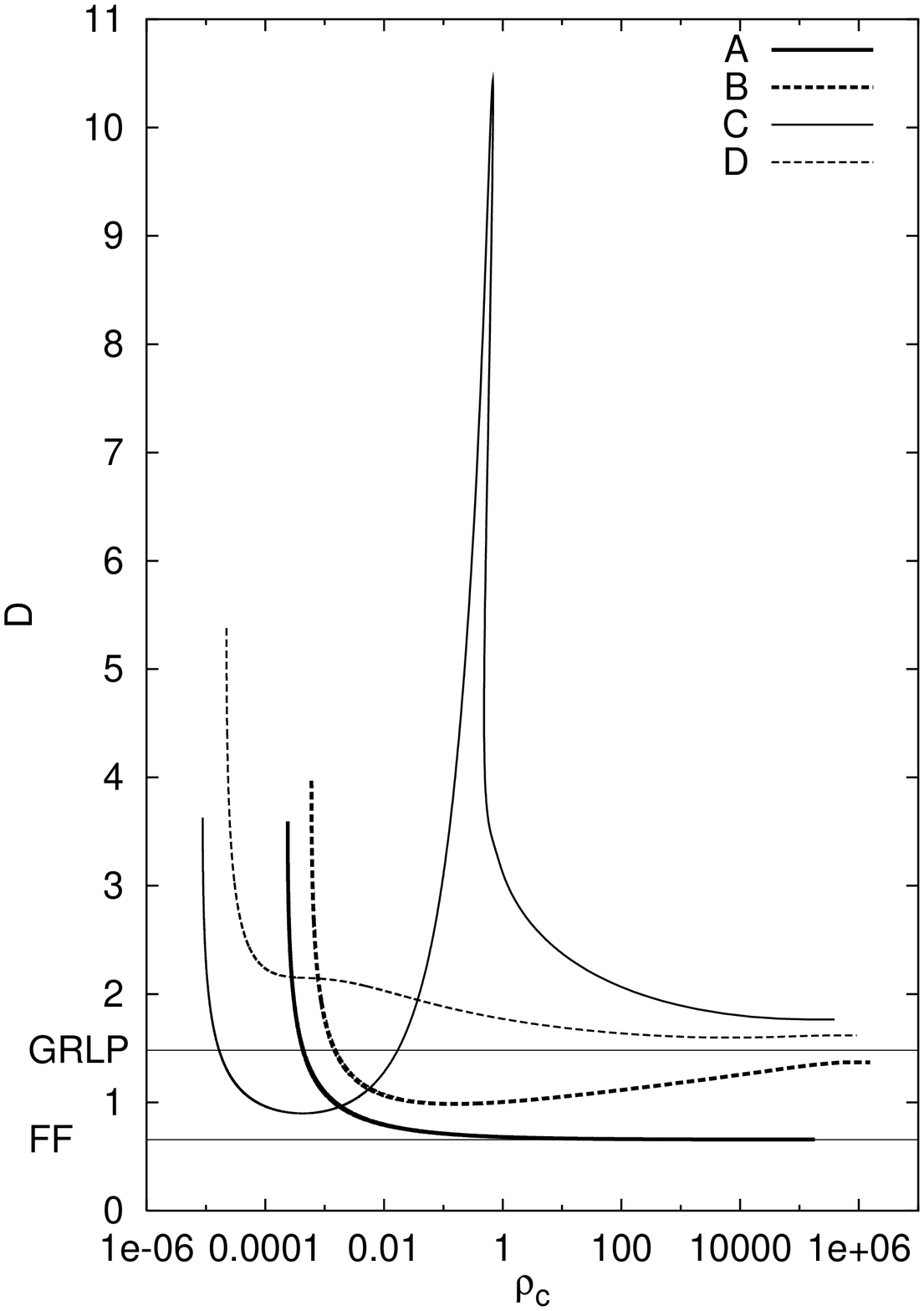}}
\centerline{(c)
\epsfysize 11cm
\epsfxsize 11cm
\epsfbox{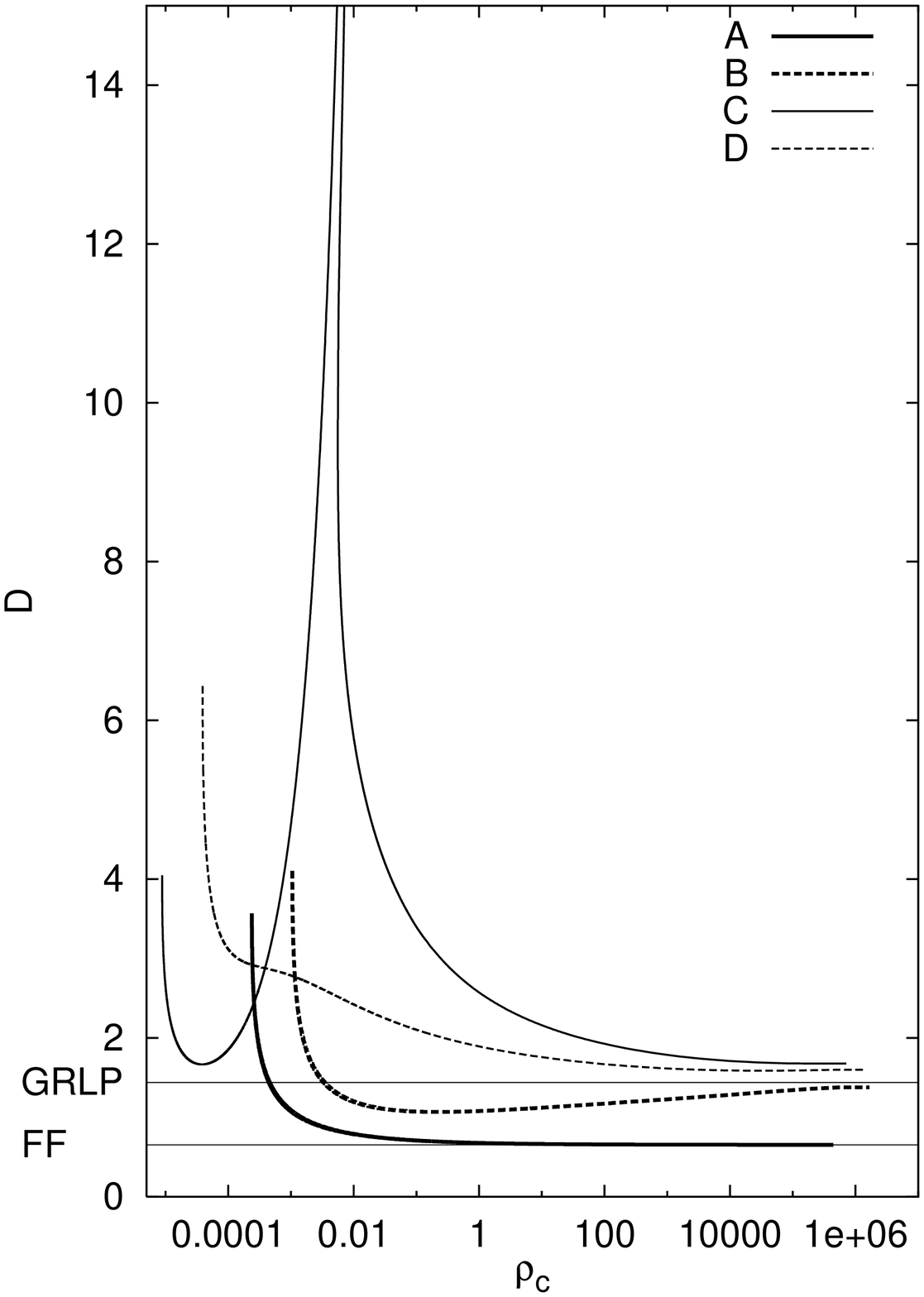}}
\centerline{(d)
\epsfysize 11cm
\epsfxsize 11cm 
\epsfbox{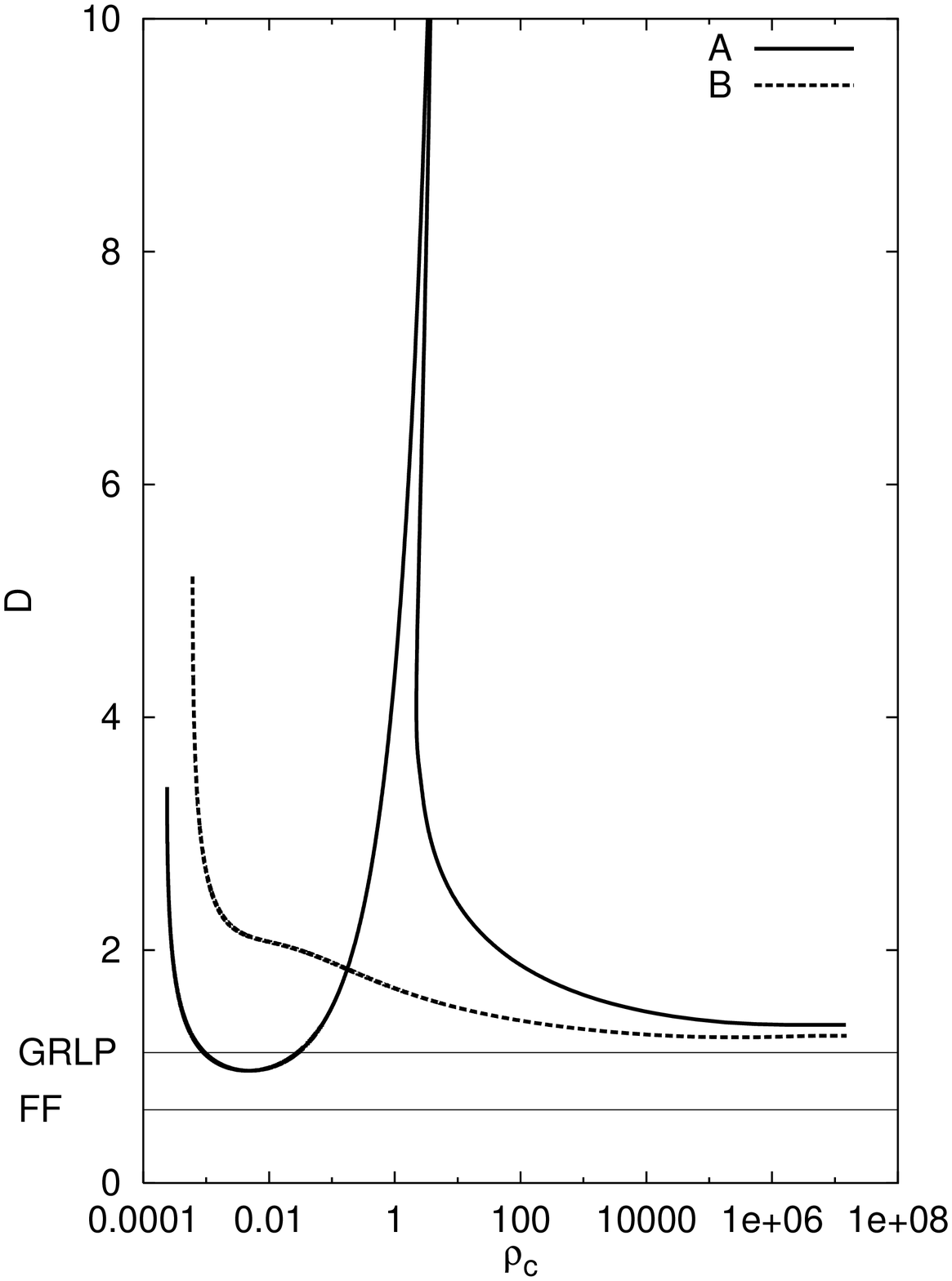}}
\caption{$D= 4\pi \rho_{c}t^{2}$ for (a) $k=0.001$, (b) $k=0.008$,
(c) $k=0.01$, and (d) $k=0.03$ are plotted.
The abscissa is the central density $\rho_{c}$.
The values of $D$ for the FF ($D=0.6653$, $0.6561$, $0.6535$ and $0.6284$)
and for the GRLP
($D=1.640$, $1.480$, $1.439$ and $1.119$) for $k=0.001$, 
$0.008$, $0.01$ and $0.03$
are also denoted, respectively.}
\label{fg:d}
\end{figure}

Moreover, in order to see which self-similar solution
the collapse approaches,
we have calculated the quantity $D$ which is defined by
\begin{equation}
  D\equiv 4\pi \rho_{c} t^{2},
\end{equation}
where $\rho_{c}$ is the central density.
This definition is consistent with Eq.~(\ref{eq:definitionD}).
Actually, we have determined the origin of $t$ by requiring
that the above defined $D$ tends to be constant.
The results for $k=0.001$, 0.008, 0.01, and 0.03 
are plotted in Figs.~\ref{fg:d}(a)--\ref{fg:d}(d).
Then, we have found that for 
$k=0.008$, 0.01, and 0.03
most models converge to the GRLP. 
Model A for $k=0.001$, 0.008, and 0.01
and model C for $k=0.001$
turn out to be trivial counterexamples 
against the convergence. We will discuss them later.
The results of numerical simulations are summarized 
in Table~\ref{tb:fate}.
The resolution of our code has not been sufficient to
show the convergence of models B and D to the GRLP 
for $k=0.001$,
although some tendency towards the GRLP has been observed. 
For dust collapse ($k=0$), we can easily find that
the above defined $D$ approaches $2/3$,
using the Lema\^{\i}tre-Tolman-Bondi solution.
Therefore, from the continuity with respect to $k$, 
it is expected that the convergence to the GRLP becomes
slower as $k$ goes to zero.
In fact, since the Newtonian approximation 
becomes good for $k\ll 1$,
it can be said that the convergence to the GRLP
for $k\ll 1$ has been confirmed by Newtonian SPH
simulations, by Tsuribe and Inutsuka~\cite{ti1999}. 
Then, we conclude that the results of numerical simulations
strongly suggest that generic spherical collapse 
converges to the GRLP in an approach to the 
singularity occurrence in both space and time.

\subsection{Interpretation}
\label{subsec:interpretation}
As we have seen in Sec.~\ref{subsec:numerical simulations},
most collapse models approach the GRLP, though
several models do not
approach the GRLP.
Here we interpret the results analytically.

First, we consider homogeneous models such as
model A for $k=0.001$, 0.008 and 0.01 and
model C for $k=0.001$. 
For an initially time symmetric homogeneous ball, 
the evolution of the central region is 
described by the closed Friedmann solution
until the rarefaction wave propagates from the surface to
the center.
The line element of the homogeneous central region is written as
\begin{equation}
ds^2=-dt^2+a^2(t)
\left[d\chi^2+\sin^2\chi (d\theta^{2}
+\sin^2\theta d\phi^{2})\right].
\end{equation}
The initial value $a_{i}$ of the scale factor $a$
and the surface value $\chi_{s}$ of the comoving coordinate $\chi$
are written using the initial density $\rho_{i}$ and the
initial circumferential radius $R_{s,i}$ as
\begin{eqnarray}
a_{i}^{2}&=&\frac{3}{8\pi\rho_{i}}, \\
\chi_{s}&=&\mbox{Arc}\sin\left(\frac{R_{s,i}}{a_{i}}\right).
\end{eqnarray}   
Therefore, the central region of the  
initially time symmetric ball
begins to contract for $k>-1/3$.
We restrict our attention to $k> 0$.
If the sound wave does not reach the center until the 
central Friedmann region collapses to the
big crunch singularity, the central region
approaches not the GRLP but the FF.
For $k \ll 1$, the central homogeneous part is well described
by the closed Friedmann solution 
with dust, which has the parametrized 
representation as
\begin{eqnarray}
a&=&a_{i}\frac{1+\cos 2\theta}{2}, \\
t-t_{i}&=&a_{i}\left(\theta+\frac{1}{2}\sin 2\theta\right). 
\end{eqnarray}
The big crunch occurs at $\theta=\pi/2$, i.e., 
$t-t_{i}=\pi a_{i}/2 = \sqrt{3\pi/(32\rho_{i})}$.
The trajectory of the rarefaction 
wave with the sound speed $c_{s}=\sqrt{k}$ 
which emanates from the surface at $t=t_{i}$
satisfies
\begin{equation}
a\frac{d\chi}{dt}=-c_{s},
\end{equation}
which can be integrated as
\begin{equation}
\chi=\chi_{s}-2 c_{s}\theta.
\end{equation}
Therefore, the condition for the sound wave not to reach the center
before the big crunch is given by
\begin{equation}
\frac{\chi_{s}}{c_{s}}>\pi.
\end{equation}
Using the free-fall velocity $v_{ff}$ defined as
\begin{equation}
v_{ff}\equiv\sqrt{\frac{2{\cal M}}{R_{s,i}}},
\end{equation}
we find the condition
\begin{equation}
\frac{c_{s}}{v_{ff}}<\frac{1}{\pi},
\label{eq:nonconvergence1}
\end{equation}
or that the compactness ${\cal M}/R_{s,i}$ satisfies 
\begin{equation}
\frac{{\cal M}}{R_{s,i}}>\frac{\pi^{2}}{2}k,
\label{eq:nonconvergence2}
\end{equation}
where we have used $k\ll 1$.
Equation (\ref{eq:nonconvergence2}) can explain the results
of the numerical simulations.
Condition (\ref{eq:nonconvergence1}) 
completely agrees with that for Newton gravity
derived by Tsuribe and Inutsuka~\cite{ti1999}, although the present 
situation can be highly relativistic.
In particular, the present analysis is valid even 
for the evolution inside an apparent horizon.
Although we have discussed the initially time symmetric case,
it is easy to derive a similar condition for the collapse
in which
the central homogeneous region
which is not swept by the rarefaction wave, is described
by the Friedmann solution, which may be 
not only the closed Friedmann solution but also
the flat or open Friedmann solution.  
The result is the same as Eq.~(\ref{eq:nonconvergence2}).

In general, we can enumerate the following trivial counterexamples
against the convergence to the GRLP.
If the central region can be initially 
described by the Friedmann solution,
then the central region does not approach
the GRLP, but the FF instead, if 
the big crunch occurs before the rarefaction wave 
reaches the center.

Moreover, it is clear that
the {\em exact} self-similar solutions other than
the GRLP do not approach the GRLP.
There are another kind of counterexamples.
We divide the initial data at $R=R_{p}$ into two regions:
the central region and the surrounding region.
If the initial data in the central region
is the same as those of the exact self-similar solutions 
other than the GRLP, and $R_{p}$ is 
so large that the sound wave cannot reach the 
center until the central singularity forms,
the collapse in the neighborhood of the center
is described not by the GRLP but by the 
self-similar solution initially prepared in the
central region.

Finally, there exists another type of counterexamples,
which can be obtained by the {\em exact fine-tuning} of parameters
which characterize the initial data.
We will discuss this type of counterexamples
in Sec.~\ref{sec:mode analysis}.

Anyway, it is clear that a set of the above trivial counterexamples
occupies only zero-measure in the space of 
the whole of regular initial data.

\section{mode analysis}
\label{sec:mode analysis}
As we have seen in Sec.~\ref{sec:numerical evidence},
the results of numerical simulations 
suggest that only the GRLP has an attractive nature.
In order to confirm this,
we examine the behavior of 
modes in linear perturbations of the self-similar solutions.

\subsection{Perturbation equations}
We consider the spherically symmetric perturbation around the 
fixed self-similar solution.
We attach the suffix $_{0}$ for the background solution.
Using the rescaling freedom, we set the arbitrary functions
$a_{\sigma}$ and $a_{\omega}$ to the background value, i.e.,
\begin{eqnarray}
a_{\sigma}&=&a_{\sigma 0}=(2D)^{\frac{2k}{1+k}}, \\
a_{\omega}&=&a_{\omega 0}=1.
\end{eqnarray}
We define the perturbation quantities as
\begin{eqnarray}
M&=&M_{0}(z)\left[1+\epsilon M_1(\tau,z)+O(\epsilon^2)\right], \\
S&=&S_{0}(z)\left[1+\epsilon S_1(\tau,z)+O(\epsilon^2)\right], \\
\eta&=&\eta_{0}(z)\left[1+\epsilon \eta_{1}(\tau,z)+O(\epsilon^2)\right], \\
y&=&y_{0}(z)\left[1+\epsilon y_1(\tau,z)+O(\epsilon^2)\right],
\end{eqnarray}
where $\epsilon$ is a small parameter which controls the expansion.
Then we find the equations for perturbations up to 
linear order of $\epsilon$ as
\begin{eqnarray}
& & M_1^{\prime}=
-\frac{k}{1+k}\frac{y_{1}}{y}
-\frac{1}{1+k}
(\dot{M_{1}}+ky^{-1}
\dot{S_{1}}), \\
& &S_1^{\prime}=\frac{1}{1+k}yy_{1}
-\frac{1}{1+k}y
(\dot{M_{1}}+ky^{-1}\dot{S_{1}}), \\
& &\frac{1}{2}(1+k)^2 y\eta^{-\frac{1-k}{1+k}}S^{-4}(M_{1}-S_{1})\nonumber \\
& &=V_{z}^{2}\left[(1-y)^2\left(\frac{k}{1+k}\eta_{1}+S_{1}\right)-
(1+k)(1-y)(\dot{S}_{1}+S_{1}^{\prime})\right]\nonumber \\
& &-\left[(k+y)^2\left(\frac{1}{1+k}\eta_{1}+3S_{1}\right)+(1+k)
(k+y)S_{1}^{\prime}\right], \\
& &y_{1}=M_{1}-\eta_{1}-3S_{1},
\end{eqnarray}
where we have omitted the suffix $_{0}$ for simplicity.

Assuming the time dependence of the perturbative 
quantities $Q_{1}(\tau,z)=e^{\lambda\tau}Q_{1}(z)$,
we find the following set of simultaneous equations:
\begin{eqnarray}
& & M_1^{\prime}=
-\frac{k}{1+k}\frac{y_{1}}{y}
-\frac{\lambda}{1+k}
(M_{1}+ky^{-1}S_{1}), \\
& &S_1^{\prime}=\frac{1}{1+k}yy_{1}
-\frac{\lambda}{1+k}y
(M_{1}+ky^{-1}S_{1}), 
\label{eq:S1prime}  \\
& &(V_{z}^{2}-k)(1-y)(k+y)y_{1}\nonumber \\
&=&
  \left[ k V_{z}^2(1-y)^2-(k+y)^2-\frac{1}{2}{\left( 1 + k \right) }^3 
	{\eta}^{-\frac{1-k}{1 + k}} 
            y S^{-4} 
\right. \nonumber \\
& & \left.
	+
     \left( 1 + k \right)  y 
        \left( V_{z}^2 \left( 1 - y \right) +(k + y) \right)  
        \lambda\right]     
   M_{1}  \nonumber \\   
& &+
\left[ \left( 1 - 2 k \right)  V_{z}^2 
              {\left( 1 - y \right) }^2  - 
           3 k {\left( k + y \right) }^2 +
           \frac{1}{2}{\left( 1 + k \right) }^3
	{\eta}^{-\frac{1 - k}{1 + k}} 
               y S^{-4} 
\right. \nonumber \\
& & \left.
	-
      \left( 1 + k \right)  
         \left( V_{z}^2 \left( 1 - y \right)  -k(k+y)
           \right)  \lambda \right]
    S_{1}.
\label{eq:Solve y1}
\end{eqnarray}
We can derive another dependent equation as
\begin{eqnarray}
& &(V_{z}^2-k)(1-y)(k+y)y_{1}^{\prime} \nonumber \\
&=&\left[kV_{z}^{2}(1-y)\left((1-y)(\ln V_{z}^{2})^{\prime}
-2y(\ln y)^{\prime}\right)
-2(k+y)y(\ln y)^{\prime}\right.\nonumber \\
& &-\frac{1}{2}(1+k)^3 \eta^{-\frac{1-k}{1+k}} y S^{-4}
\left(-\frac{1-k}{1+k}(\ln \eta)^{\prime}
+(\ln y)^{\prime}-4 (\ln S)^{\prime}\right) \nonumber \\ 
& &\left.+(1+k)y \left(
\left(V_{z}^{2}(1-2y)+(k+2y)\right) (\ln y)^{\prime}
+V_{z}^{2}(1-y)(\ln V_{z}^{2})^{\prime}\right)
\lambda \right] M_{1} \nonumber \\
& &+\left[ k V_{z}^2(1-y)^2-(k+y)^2-\frac{1}{2}{\left( 1 + k \right) }^3 
	{\eta}^{-\frac{1-k}{1 + k}} 
            y S^{-4} 
\right. \nonumber \\
& & \left.
	+
     \left( 1 + k \right)  y 
        \left( V_{z}^2 \left( 1 - y \right) +(k + y) \right)  
        \lambda\right]     
   M_{1}^{\prime} \nonumber \\
& &+\left[(1-2k)V_{z}^{2}(1-y)\left((1-y)(\ln V_{z}^{2})^{\prime}
-2y(\ln y)^{\prime}\right)
-6k(k+y)y(\ln y)^{\prime} \right.\nonumber \\
& &+\frac{1}{2}(1+k)^{3}\eta^{-\frac{1-k}{1+k}}y S^{-4}
\left(-\frac{1-k}{1+k}(\ln \eta)^{\prime}
+(\ln y)^{\prime}-4(\ln S)^{\prime}\right)\nonumber \\
& &\left.-(1+k)\left(V_{z}^{2}(1-y)(\ln V_{z}^{2})^{\prime}
-(V_{z}^{2}+k)y(\ln y)^{\prime}\right)
\lambda\right]S_{1} \nonumber \\
& &+\left[ \left( 1 - 2 k \right)  V_{z}^2 
              {\left( 1 - y \right) }^2  - 
           3 k {\left( k + y \right) }^2 +
           \frac{1}{2}{\left( 1 + k \right) }^3
	{\eta}^{-\frac{1 - k}{1 + k}} 
               y S^{-4} 
\right. \nonumber \\
& & \left.
	-
      \left( 1 + k \right)  
         \left( V_{z}^2 \left( 1 - y \right)  -k(k+y)
           \right)  \lambda \right]
    S_{1}^{\prime} \nonumber \\
& &-\left[V_{z}^{2}(1-y)(k+y)(\ln V_{z}^{2})^{\prime}
+(V_{z}^{2}-k)(1-k-2y)y(\ln y)^{\prime}\right] y_{1}.
\label{eq:y1prime}
\end{eqnarray}
where $(\ln y)^{\prime}$ and $(\ln V_{z}^{2})^{\prime}$ are given by
\begin{eqnarray}
(\ln y)^{\prime}&=&(\ln M)^{\prime}-(\ln \eta)^{\prime}-3(\ln S)^{\prime}, \\
(\ln V_{z}^{2})^{\prime}&=&2\frac{1-k}{1+k}-2\frac{1-k}{1+k}(\ln \eta)^{\prime}
-4 (\ln S)^{\prime}.
\end{eqnarray}

Then, we examine boundary conditions which the perturbation
quantities should satisfy at the boundaries.
First, we consider the regular center $z=+0$.
At the regular center, the definition of $y$ implies that 
the perturbation of $y$ must vanish at $z=+0$, since 
the background solution already satisfies the boundary condition
in full order. Then we obtain
\begin{equation}
y_{1}=0.
\end{equation}
at $z=+0$. From Eq.~(\ref{eq:Solve y1}), it implies the following
condition:
\begin{equation}
M_{1}+3k S_{1}=0.
\end{equation}
Then, the perturbation solutions which are regular at the 
center for fixed $\lambda$ 
are parametrized by one parameter $\Delta$.
The boundary condition at $z=+0$ is written as
\begin{eqnarray}
y_{1}&=&0,\\
M_{1}&=&\frac{k}{1+k}\Delta,\\
S_{1}&=&-\frac{1}{3(1+k)}\Delta,
\end{eqnarray}
where 
\begin{equation}
\Delta\equiv\eta_{1}(0)=\frac{\delta \rho}{\rho}(t=-1,r=0).
\end{equation}
$\Delta$ only scales $y_{1}$, $M_{1}$, and $S_{1}$
because we are only considering the linear perturbations.
Hence,
we can set the parameter $\Delta$ as $\Delta=1$
without loss of generality.

Next, we consider the sonic point $z=z_{sp}$.
At the sonic point, we require that the density 
perturbation is regular. 
It implies that $M_{1}$, $S_{1}$ and $y_{1}$ 
must satisfy the condition
that the right-hand side of Eq. (\ref{eq:y1prime})
vanishes at the sonic point.
Only for a discrete set of $\lambda$, there exists 
a solution of perturbation equations that is regular
both at the regular center and at 
the sonic point.
Thus we can obtain eigenvalues $\lambda$ and 
the associated eigenmodes.

\subsection{Results of mode analysis}
It is found that the system has a gauge mode
with the eigenvalue $\lambda$ given by
\begin{equation}
\lambda=\frac{1-k}{1+k}.
\end{equation}
The mode functions are given by
\begin{eqnarray}
M_{1}&=&\frac{\Delta}{2}(\ln M)^{\prime},\\
S_{1}&=&\frac{\Delta}{2}(\ln S)^{\prime},\\
y_{1}&=&\frac{\Delta}{2}(\ln y)^{\prime}.
\end{eqnarray}
This mode corresponds to the following gauge transformation:
\begin{eqnarray}
(-t) &\to& (-t)-\epsilon \frac{\Delta}{2} (-t)^{\frac{2k}{1+k}},\\
r &\to& r 
\end{eqnarray}
or, equivalently,
\begin{eqnarray}
\tau &\to & \tau -\epsilon \frac{\Delta}{2}e^{\frac{1-k}{1+k}\tau}, \\
z &\to& z + \epsilon \frac{\Delta}{2} e^{\frac{1-k}{1+k}\tau}z.
\end{eqnarray}

The eigenvalues of physical repulsive modes for $k=0.001$, $0.008$, 
$0.01$, and $0.03$ are summarized 
in Table~\ref{tb:modes}, where $\lambda\in \mbox{\bf R}$ is assumed.
For the FF and GRLP, there exists no repulsive mode.
On the other hand, the GRHA and GRHB have one and two
repulsive modes, respectively.
Therefore, it is found that only the FF and GRLP 
can describe the final stage of the central region
of generic collapse.
Together with the existence of the ``kink" instability in 
the FF and the self-similar solutions 
which are not analytic at the sonic point~\cite{op1990,op1988b}, 
we conclude that the 
GRLP is the only self-similar solution that can be an attractor.
For the GRHA, there exists only one repulsive mode.
This solution corresponds to the critical solution in the black hole 
critical behavior.
Only when one parameter $p$, which parametrizes
initial data, is fine-tuned around the critical value $p^{*}$
for the black hole formation,
this solution has importance as a critical solution.
The critical exponent $\gamma$, 
which appears in the scaling law
of the formed black hole mass 
$M_{BH}\propto (p-p^{*})^{\gamma}$, 
is given by the inverse of this 
lowest repulsive mode.
For $k=0.01$,
the eigenvalue we have obtained agrees well
with Maison~\cite{maison1996}.
Since the GRHA solution has a repulsive mode, 
it is not relevant for the behavior of generic collapse.
In particular, the final stage of the collapse
can be described by the GRHA if the parameter is
{\em exactly} fine-tuned, i.e., $p=p^{*}$.
The GRHB has two repulsive modes.
It is expected from the mode analyses in Newton gravity~\cite{hn1997},
that the solution with $n$ oscillations has $n$ repulsive modes.
In order for the solution with $n$
oscillations to be relevant, $n$ parameters must be fine-tuned.
If the fine-tuning is not exact, the perturbation grows 
into nonlinear regime. Then, it is expected that 
the collapse will approach the 
GRLP or disperse away.

\section{discussions}

First, we discuss the validity of self-similarity hypothesis.
The results of the numerical simulations and mode analysis
strongly suggest that 
generic spherical collapse of a perfect fluid
with small $k$ converges to the GRLP in an approach to
the singularity. This means that the GRLP is an attractor solution.
Moreover, in Sec.~\ref{subsec:interpretation}, we have discussed
several counterexamples against the convergence to the GRLP.
It is surprising that these counterexamples are 
exactly self-similar in the neighborhood of the center 
or at least asymptotically self-similar.
It should be noted that non-flat Friedmann solution
also approaches the FF asymptotically
in an approach to the big crunch.
Therefore, we can conjecture that 
{\em any cloud of a perfect 
fluid collapses in a self-similar manner 
in an approach to the singularity.}

Next we discuss the implications of the convergence to the GRLP
in the context of the cosmic censorship.
The cosmic censorship conjecture states that 
a naked singularity does not form in the gravitational
collapse which develops from generic initial data
with matter fields which obey a physically reasonable
equation of state. 
For spherical collapse, the convergence to the GRLP, 
which we have observed in this paper, 
means that
a naked singularity forms in generic collapse 
for an equation of state $P=k\rho$
for $0<k\alt 0.0105$, because the GRLP has a naked singularity
for that range of $k$~\cite{op1990,op1987,op1988a}.

Here we should give the precise terminology of a naked singularity.
In this article, we refer to a singularity that 
can be seen by some observer 
as a naked singularity.
In contrast, a naked singularity which can be seen 
from infinity
is called a globally naked singularity.
Whether a naked singularity is globally naked 
is determined not only by the central region but also
by the surrounding region.
In fact, a naked singularity
can be globally naked through the matching of the central region
with an appropriate surrounding region.

Now that we have the precise terminology, we can discuss the 
consistency of our results with previous works 
on the black hole critical behavior.
At first sight, our results seem to be inconsistent with 
the formation of an apparent horizon observed in 
numerical simulations showing the
black hole critical behavior.
In fact, this is not the case.
Since the convergence is only for the neighborhood
of the center, we cannot say whether the formed naked singularity 
is locally naked or globally naked.
Because the formation of an apparent horizon
only implies the existence of an event horizon
outside or coinciding with it,
it does not exclude the formation of locally naked singularity
at the center.

If the cosmic censorship is true, then there are three possibilities.
One is that deviations from spherical symmetry may play a crucial role
in the nakedness of the formed singularity.
Although
there has been no systematic study on the effect 
of violation of spherical symmetry
in inhomogeneous gravitational collapse, 
Shapiro and Teukolsky~\cite{st1991} reported some numerical results
that suggest the occurrence of naked singularity in 
the axisymmetric collapse of collisionless particles.
In contrast, 
Iguchi et al.~\cite{inh1998,ihn1999,ihn2000}
and Nakao et al.~\cite{nhi2000}
reported some kind of instability along the Cauchy horizon
associated with a globally naked singularity.
The second possibility is that the small value of $k$ is not 
allowed for extremely high-density matter fields.
However, it seems to be strange that the consistency of
classical theory of gravity restricts the equation of 
state for high-density matter fields, 
which is determined by a collection of 
various microscopic physics.
The third possibility is that the fluid description 
for high-density matter may be responsible.

Whether or not the cosmic censorship conjecture
is true, the convergence 
to the GRLP strongly suggests
that there can appear an extremely high-density or high-curvature
region which can be seen by an observer.
Even for such an ``approximate" naked singularity,
it has been shown that explosive radiation is 
emitted due to quantum 
effects~\cite{bsvw1998a,bsvw1998b,vw1998,hin2000a,hin2000b,hisntvw2000,ts2000,ih2000}.
Furthermore, in a practical sense, if the curvature scale reaches 
the Planck scale, it should be regarded as a singularity because 
it is considered beyond the scope of classical general relativity.

Self-similar solutions we have obtained here
approach those in Newton gravity in the limit $k\to 0$.
Therefore, 
the important consequence is that critical phenomena
associated with the Hunter (a) Newtonian self-similar solution
should be observed in the collapse of isothermal gas in Newton gravity.
These critical phenomena will be very similar to the critical
phenomena in the black hole formation in general relativity.
Only one parameter $p$ has to be fine-tuned 
closely to the critical value $p^{*}$. 
In particular, some order parameter ${\cal A}$ follows 
the scaling law ${\cal A} \propto (p-p^{*})^{\gamma}$
in the near critical regime,
where $\gamma$ is given by the inverse of the eigenvalue 
of the only one repulsive mode of the Hunter (a) solution.
Unfortunately, the eigenvalue of the repulsive mode
of the Hunter (a) solution has not been known yet.
However, by extrapolating our results on the GRHA
to the limit $k\to 0$,
we can predict that the critical exponent $\gamma$ is given by
$\gamma\approx 0.11$.
The candidate for the order parameter ${\cal A}$
will be, for example, the mass of the initially formed core,
if we assume the realistic equation of state 
for dense gas.

\section{conclusions}
The results of the numerical simulations and 
mode analysis strongly suggest that
the general relativistic Larson-Penston solution
is an attractor solution of spherically symmetric 
gravitational collapse of a perfect fluid
with an adiabatic equation of state $P=k\rho$
for $0<k\alt 0.036$ in general relativity.
Since a naked singularity forms
in the general relativistic Larson-Penston
solution for $0<k\alt 0.0105$,
the analysis in this paper means
the violation of cosmic censorship in spherically 
symmetric case.
This will be the strongest known counterexample against
the cosmic censorship ever.
This also provides a strong evidence for the self-similarity hypothesis
in general relativistic gravitational collapse.

\acknowledgments
We are very grateful to B.J.~Carr, H.~Kodama, T.P.~Singh, 
K.~Nakao, T.~Hanawa, and S.~Inutsuka for helpful discussions and comments.
We would also like to thank K.~Maeda for continuous encouragement. 
This work was partly supported by the 
Grant-in-Aid for Scientific Research (No. 05540)
from the Japanese Ministry of
Education, Science, Sports and Culture.

\appendix

\newpage
\begin{table}[htbp]
\begin{center}
  \caption{Values of $D$ for self-similar solutions.}
  \label{tb:D}
    \begin{tabular}{c|cccc} 
        & $k=0.001$ & 0.008 & 0.01 & 0.03 \\ \hline
      FF  & $2/(3\times 1.001^{2})$ & 
      $2/(3\times 1.008^{2})$ & $2/(3\times 1.01^{2})$ 
	&$2/(3\times 1.03^{2})$ \\
      GRLP  & 1.640 & 1.480 & 1.439 & 1.119 \\
      GRHA  & $1.675\times 10^{3}$ & 
      $1.345\times 10^{3}$  & $1.265\times 10^{3}$& $7.204\times 10^{2}$\\
      GRHB  & $7.170\times 10^{4}$ & $4.903\times 10^{4}$ 
      & $4.414\times 10^{4}$ & $1.650\times 10^{4}$ \\
    \end{tabular}
  \end{center}
\end{table}
\begin{table}[htbp]
\begin{center}
  \caption{Models for numerical simulations.}
  \label{tb:models}
    \begin{tabular}{c|cc} 
      Models  & Initial Density Profile & Initial Compactness (${\cal M}/R_{s,i}$) \\ \hline
      A  & Homogeneous & $1/10$ \\
      B  & Inhomogeneous & $1/10$ \\
      C  & Homogeneous  & $1/30$ \\
      D  & Inhomogeneous & $1/30$ \\
    \end{tabular}
  \end{center}
\end{table}
\begin{table}[htbp]
\begin{center}
  \caption{Asymptotic behavior of the collapse models.}
  \label{tb:fate}
    \begin{tabular}{c|cccc} 
      Models  & $k=0.001 $& 0.008 & 0.01 & 0.03  \\ \hline
      A    & FF  & FF    & FF   & GRLP   \\
      B    & GRLP? & GRLP   & GRLP  & GRLP   \\
      C    & FF & GRLP   &  GRLP & (Dispersion)\\
      D    & GRLP? & GRLP   & GRLP & (Dispersion)\\
    \end{tabular}
  \end{center}
\end{table}

\begin{table}[htbp]
\begin{center}
  \caption{Eigenvalues $\lambda$ for repulsive modes 
	of self-similar solutions.}
  \label{tb:modes}
    \begin{tabular}{l|cccc} 
        & $k=0.001$ & 0.008 & 0.01 & 0.03
      \\ \hline
	 FF  & None & None   & None    & None    \\
	 GRLP& None &None   & None    & None    \\
         GRHA& 9.39 & 8.88   & 8.75    & 7.62    \\
	 GRHB& 5.43 & 5.10   & 5.02    & 4.27    \\
	     & $5.62 \times 10$ & $4.90\times 10$ & $4.71\times 10$ & $3.30\times 10$ \\
    \end{tabular}
  \end{center}
\end{table}

\newpage
\newpage
\newpage
\newpage

\end{document}